\documentclass[aoas]{imsart}

\RequirePackage{amsthm,amsmath,amsfonts,amssymb}
\RequirePackage[authoryear]{natbib}
\RequirePackage[colorlinks,citecolor=blue,urlcolor=blue]{hyperref}
\RequirePackage{graphicx}
\usepackage[ruled,vlined]{algorithm2e}

\startlocaldefs
\theoremstyle{plain}

\newtheorem{theorem}{Theorem}[section]

\theoremstyle{remark}
\newtheorem{definition}[theorem]{Definition}
\newtheorem*{example}{Example}

\newcommand{\Id}[2]{{\mbox{Id}}_{#1}(#2)} 
\DeclareMathOperator*{\mean}{mean}

\newcommand{\Y}{\mathbf{Y}} 
\newcommand{\W}{\mathbf{W}}
\newcommand{\WC}[1][]{\mathbf{W}^{{\tiny \mbox{comp}} #1}} 
\newcommand{\WT}[1][]{W^{{\tiny \mbox{tot}} #1}} 
\newcommand{\WTS}{\mathcal{W}^{\tiny \mbox{tot}}} 
\newcommand{\thetaT}{\theta^{\tiny \mbox{tot}}} 
\newcommand{\thetaC}{\theta^{\tiny \mbox{comp}}} 
\newcommand{\set}[1]{\mathcal{#1}} 

\endlocaldefs

\begin{document}

\begin{frontmatter}
\title{Scale Reliant Inference}
\runtitle{Scale Reliant Inference}

\begin{aug}
\author[A,B]{\fnms{Michelle Pistner}~\snm{Nixon}},
\author[C]{\fnms{Kyle C.}~\snm{McGovern}},
\author[D]{\fnms{Jeffrey}~\snm{Letourneau}},
\author[D]{\fnms{Lawrence A.}~\snm{David}},
\author[E,F]{\fnms{Nicole A.}~\snm{Lazar}},
\author[G,H,I]{\fnms{Sayan}~\snm{Mukherjee}},
\and \\
\author[A,B,C,E,J]{\fnms{Justin D.}~\snm{Silverman}\ead[label=e1]{JustinSilverman@psu.edu}\orcid{0000-0002-3063-2098}},
\address[A]{Denotes Equal Contribution}

\address[B]{College of Information Sciences and Technology, The Pennsylvania State University\printead[presep={,\ }]{e1}}
\address[C]{Program in Bioinformatics and Genomics, The Pennsylvania State University}
\address[D]{Department of Molecular Genetics and Microbiology, Duke University}
\address[E]{Department of Statistics, The Pennsylvania State University}
\address[F]{Huck Institutes for the Life Sciences, The Pennsylvania State University}
\address[G]{Departments of Statistical Science, Mathematics, Computer Science, Biostatistics \& Bioinformatics, Duke University}
\address[H]{Center for Scalable Data Analytics and Artificial Intelligence, University of Leipzig}
\address[I]{Max Planck Institute for Mathematics in the Natural Sciences}
\address[J]{Department of Medicine, The Pennsylvania State University}
\end{aug}

\begin{abstract}
Many scientific fields, including human gut microbiome science, collect multivariate count data where the sum of the counts is unrelated to the scale of the underlying system being measured (e.g., total microbial load in a subject's colon). This disconnect complicates downstream analyses such as differential analysis in case-control studies. This article is motivated by a novel study of \textit{in vitro} human gut microbiome models. Popular tools for analyzing these data led to dramatically elevated rates of both false positives and false negatives. To understand those failures, we  provide a formal problem statement that frames these challenges of scale in terms of the classical theory of identifiability. We call this the problem of Scale Reliant Inference (SRI). We use this formulation to prove fundamental limits on SRI in terms of criteria such as consistency and type-I error control. We show that the failures of existing methods stem from a fundamental failure to properly quantify uncertainty in the system scale. We demonstrate that a particular type of Bayesian model called a Bayesian Partially Identified Model (PIMs) can correctly quantify uncertainty in SRI. We introduce Scale Simulation Random Variables (SSRVs) as a flexible and efficient approach to specifying and inferring Bayesian PIMs. In the context of both real and simulated data, we find SSRVs drastically decrease type-I and type-II error rates. 
\end{abstract}


\end{frontmatter}

\section{Introduction}\label{sec:intro}

  This article is motivated by statistical challenges that arise in many areas of science and have become a central focus of modern microbiome research. Microbiome sequence count data are typically represented as an \(D\times N\) table of counts \(\Y\) with elements \(\Y_{dn}\)  representing the number of DNA sequences observed that map to the \(d\)-th microbial taxon in the \(n\)-th biological sample (e.g., a fecal sample from the \(n\)-th study participant). The measurement process is non-trivial yet, for the purposes of the present article, can be thought of as a multivariate sample: there is a large pool of microbes in the human gut that could be measured, and our measurements capture a small, arbitrary-sized sample of that pool~\citep{silverman2018dlm,roche2022composition}. By arbitrary sized, we allude to the fact that the size of the sample (the sequencing depth \(\sum_{d}\Y_{dn}\)) is typically independent of the \textit{scale} of the biological system from which that sample was obtained (e.g., the total number of microbes in the subject's colon)~\citep{lovell2011proportions,props2017absolute,vandeputte2017quantitative}. As a result, many authors call these data \textit{compositional}, alluding to the idea that the data only provides information about the proportional abundance of each microbe and not their absolute abundance~\citep{gloor2016allrelative,lovell2011proportions,gloor2017notoptional}. The problem is that many of the scientific questions asked of these data require information about absolute abundances rather than just proportional abundances.

  For concreteness, this article introduces a novel microbiome study that aimed to quantify the limits of \textit{in vitro} artificial gut vessels as models of human gut microbiota. Eight artificial gut systems were inoculated with a human fecal microbial community sample. The inoculated vessels incubated those communities for two weeks.   Our goal was to identify which of the \(D\) microbial taxa in the community can grow within the vessels, quantified as an increased abundance on day 14 compared to day 1. This is an example of a \textit{differential abundance analysis} which aims to identify taxa that differ in abundance between two experimental conditions. For clarity, we are not interested in changes in proportional abundances as, for example, a taxon may decrease in relative abundance between days 1 and 14 simply because its growth was outpaced by some other microbe that grew more rapidly. While differential abundance analyses are performed in many, if not most, microbiome studies~\citep{lin2020analysis}, we find that popular tools for the task lead to exceedingly high rates of false positives and false negatives in our study. In this article, we argue that this is not simply an issue of data size, measurement error, or choice of parametric model but a fundamental failure of these methods to properly quantify uncertainty originating from the lack of scale information in these data.

  The main contributions of this article are the following. First, we provide a formal problem statement that defines what it means to ask a question that depends on scale using data that lacks information about scale. We call this the problem of Scale Reliant Inference (SRI) and frame it in terms of the classical theory of identifiability. We discuss the advantages of our formulation over the related problem of \textit{Compositional Data Analysis} (CoDA). Second, we use this formulation to prove fundamental limits on SRI. Contrary to numerous published claims, we prove that classical inferential criteria such as consistency and Type-I error control cannot be attained in many SRI problems. In contrast, we show that Bayesian inferential criteria can be achieved by a specialized class of models called Bayesian Partially Identified Models (PIMs). We introduce a simulation framework called Scale Simulation Random Variables (SSRVs), which can simplify the specification and inference of Bayesian PIMs for SRI. Finally, we demonstrate SSRVs through both simulated and real data analysis; in particular, we use SSRVs to tackle the differential abundance
  analysis at the core of the artificial gut vessel study introduced above. In the context of real data, we find SSRVs control false discovery rates while popular alternatives display rates as high as 30 percent. Moreover, SSRVs drastically reduce false rates, identifying over three times more true positives than alternatives. Our methods have already been implemented into popular software tools~\citep{nixon2024aldex2}. Moreover,
  Our findings, namely the dramatic decreases in Type-I and/or Type-II error rates compared to existing tools, has been validated in the context of other microbiome datasets~\citep{nixon2024aldex2} and meta-transcriptomics~\citep{dossantos2024aldexvaginal}. 

  While this article focuses on applications to microbiome analysis, the problems discussed and the theory and methods developed are more general. Beyond microbiome studies, scale reliant inference poses problems in other areas of biomedicine (e.g., gene expression studies)~\citep{fernandes2014unifying,lovell2011proportions}, macro-ecology~\citep{billheimer2001statistical,jackson1997compositional,murphy2016capture}, political poling~\citep{topirceanu2020polling}, demography~\citep{rossi1987urban}, and epidemiology~\citep{silverman2020ili}. In all these fields, multivariate count data is collected where the total counts are arbitrary and unrelated to the measured systems' underlying scale (i.e., size).

\section{A Theory for Scale Reliant Inference}

In this section we introduce a theory for scale reliant inference from data that lacks information about scale. We also discuss implications of this theory on inference procedures. First we set notation that will be used throughout the paper.

\subsection{Notation}
\label{sec:notation}

We use \(P_{a,b}\) to denote a probability model which is a function mapping
from a parameter space \(\set{A}\times \set{B}\) to a set of distributions
\(\set{P}\). For a random variable \(x\)  the notation \(P_{a,b}(x)\) denotes a specific distribution of \(x\) within the set \(\set{P}\), a point within the image of the model
\(P_{a,b}\). The identification region  $\Id{a,b} P$
is the pre-image of the
model at a specific distribution \(P\in \set{P}\): \[\Id{a,b} P=\{(a,b)\in
\set{A}\times \set{B}: P_{a,b}=P\}.\] The model is identified at \(P\) if
\(\Id{a,b}P\) is a singleton set. We define the marginal identification region
of \(a\) as \[\Id{a}P=\{a\in \set{A}: \exists \: b\in \set{B},
P_{a,b}=P\}.\] The marginal identification region of \(b\) (\(\Id{b}P\)) is defined
analogously. Statements about the identifiability of a single parameter within a multi-parameter model refer to the marginal identification
region: i.e., \(a\) is identified at \(P\) if \(\Id{a}P\) is a singleton set. While
identification can depend on the specific point \(P\in
\set{P}\), for brevity, we will often omit this detail when the specific
point can be inferred from context.

We consider a \(D \times N\) matrix of observed data \(\Y\) which is a
measurement of a \(D \times N\) dimensional underlying system \(\W\). We place
no mathematical restrictions on \(\Y\) and only require that \(\W\) is positive-valued (integer or continuous) such that the elements \(\W_{dn}\) represent the amount of class \(d\) in source \(n\) (e.g., the abundance of taxon \(d\) in
subject \(n\)'s gut). The system \(\W\) can be uniquely decomposed into its
composition (\(\WC\), proportions) and scale (\(\WT\), totals) via:
\begin{align}
\mathbf{W}_{dn} &= \WC_{dn} \, \, \WT_{n} \label{eq:comptot} \\ 
\WT_{n} &= \sum_{d=1}^{D}\mathbf{W}_{dn}. \nonumber
\end{align}
This implies that \(\WC\) is a \(D\times N\) matrix with columns that sum to
\(1\), and \(\WT\) is a positive-valued \(N\)-vector. We denote rows of a matrix with a placeholder dot, e.g., \(\W_{\cdot n}\) is the \(n\)th row of \(\W\). 

\subsection{Formal Problem Statement}\label{sec:problem-formulation}

This subsection presents a formal definition of Scale Reliant Inference (SRI)
followed by an example of SRI in microbiome data analyses. 

Assume the observed data \(\Y\) are generated following a probability model
\(P_{\WC,\WT}\). We say that \(\Y\) lacks information about scale if the
parameter \(\WT\) is unidentified at \(P_{\WC, \WT}(\Y)\). For simplicity, this
article assumes that \(\WC\) is identified. We make this assumption to focus our
attention on the impact of scale and not other sources of measurement
limitations associated with \(\WC\) (such as PCR bias in microbiome
studies~\cite{silverman2021measuring}). We denote data generation models where
\(\WC\) can be identified but \(\WT\) cannot as \textit{compositional surveys}.
We choose this name to highlight the connection to compositional data. That
connection, along with important distinctions between SRI and compositional data
analysis are reviewed in Section~\ref{sec:coda-comparision}.

SRI can be framed as learning the map $\Y \mapsto \theta$ where
$\theta := \theta\left(\WC, \WT\right)\in \Theta$. We call \(\theta\) a \textit{target
  estimand} because it is the quantity one wants to estimate. We also introduce
a nuisance function \(\lambda=\lambda(\WC, \WT)\) to ensure that the following
concatenated function \(f\) defined as
\[(\theta, \lambda)=f(\WC, \WT)=(\theta(\WC,
  \WT), \lambda(\WC, \WT))\] is invertible. With this
condition, the model \(P_{\WC, \WT}\) can be reparameterized in
terms of \((\theta, \lambda)\) such that
\(P_{\WC, \WT}=P_{f^{-1}(\theta, \lambda)}\). We define SRI in terms of the identifiability of \(\theta\) within this
reparameterized model.

\begin{definition}[Scale Reliant and Scale Invariant Inference]
  If \(\Y\) is a compositional survey, the problem of estimating \(\theta\) is
  \textit{Scale Reliant Inference} (SRI) if \(\theta\) is unidentified at
  \(P_{\theta, \lambda}(\Y)\). Otherwise (if \(\theta\) is identified), the
  estimation problem is \textit{Scale Invariant Inference} (SII).
\end{definition}

To help motivate this definition, consider that, if \(\Y\) is a compositional
survey, the only way that \(\theta\) can be unidentified is if it depends on
\(\WT\) (since \(\Y\) provides information about \(\WC\) and
\(\theta\) is a known function). Therefore, our problem statement provides focus
to our study: it restricts attention to the setting where the lack of
information about \(\WT\) contained in \(\Y\) impedes our ability to
uniquely identify \(\theta\).

\begin{example}[Differential Abundance Analysis from Microbiome Data]
  The analysis of human gut microbiome data often centers around a \(D\times N\)
  sequence count table \(\Y\) with elements \(\Y_{dn}\) denoting the number of
  sequenced (measured) DNA molecules that map to a particular microbial taxon
  \(d\) in fecal sample \(n\). For concreteness, we provide a toy dataset
  (\(D=4\)) in Supplementary Section A. The measurement process is complex and
  requires many experimental and computational steps to transform fecal samples
  into a count table \(\Y\)~\citep{caporaso2011global}. We model the measurement
  process as random, there exist some true system with composition \(\WC\) and
  scale \(\WT\) but it is only observed through the noisy measurements \(\Y\).
  While the true number of microbes in a system is integer valued; for
  simplicity, we assume that those counts are large enough that we can
  approximate \(W_{dn} \in \mathbb{R^{+}}\). 

Many authors call these data compositional owing to the measurement process. The measurement process can be thought of as a type of poll: there is a pool of microbes that could be measured, but only an arbitrarily sized subset of those are~\citep{fernandes2014unifying,lovell2011proportions}. Moreover, due to complexities such as DNA amplification, it is not even possible to assume that the system has more sequences than the sample (\(\W_{dn}>\Y_{dn}\)). Overall, the scale of the underlying system is lost through this process leading to data that provides information about the system composition but not the system scale~\citep{vandeputte2017quantitative,props2017absolute}. We model sequence count data as a compositional survey to reflect this feature of the data.

The goal of many microbiome studies is to determine which, if any, taxa differ in abundance between two mutually exclusive conditions (e.g., disease versus health). Letting \(x_{n}\in \{0, 1\}\) denote the condition of sample \(n\), these \textit{differential abundance analyses}, often center on estimation of Log-Fold-Changes (LFCs):
\begin{equation}
  \label{eq:lfc}
  \theta_{d} = \mean_{n:x_{n}=1}\log \W_{dn}-\mean_{n:x_{n}=0}\log \W_{dn}. 
\end{equation}
LFCs are therefore the target estimand in differential abundance analyses. 
Combining Equations~\eqref{eq:comptot} and \eqref{eq:lfc} leads to the following relationship
\begin{align}
\theta_{d}&=\underbrace{\mean_{n:x_{n}=1}(\log \WC_{dn}) -\mean_{n:x_{n}=0}(\log \WC_{dn})}_{\thetaC_{d}}+ \underbrace{\mean_{n:x_{n}=1}(\log \WT_{n}) -\mean_{n:x_{n}=0}(\log \WT_{n})}_{\thetaT} \nonumber  \\ 
 &= \thetaC_{d} + \thetaT. \label{eq:lfc-decomposition}
\end{align}
In words, the LFC estimand requires knowledge of how the proportional abundance of the \(d\)-th taxon changes between conditions (\(\thetaC_{d}\)) as well as how total microbial load changes between conditions (\(\thetaT\)). Because microbiome data (\(\Y\)) is a compositional survey, it follows that  \(\thetaT\), and therefore \(\theta_{d}\), are not identifiable. By definition, differential abundance analysis from microbiome data is a problem of Scale Reliant Inference. 

Before progressing, we discuss a subtlety of our model that is a source of frequent confusion. 
There are often two sources of uncertainty in SRI. Using LFC estimation from microbiome data as an example, there will will be uncertainty in \(\WT\), and therefore \(\thetaT\), since both are unidentified. Yet we will also have uncertainty in \(\WC\), and therefore \(\thetaC\), since we only observe a noisy, finite sample \(\Y\) and not \(\WC\) directly. To clarify, our assumption that \(\WC\) is identifiable does not imply that \(\WC\) is known. 
Identifiability is a feature of the distribution of observables \(P_{\WT, \WC}(\Y)\); yet, we cannot know that distribution without potentially infinite number of replicate studies. In microbiome data analyses, we can have substantial uncertainty in \(\WC\) due to the variability in the count data \(\Y\).  Still, this article places comparatively less emphasis on this later source of compositional uncertainty as there is already substantial literature on that subject which we will use in later sections (e.g., \cite{gloor2016uncertainty,silverman2018dlm}). 
\end{example}

\subsection{Comparing Compositional Data Analysis and SRI}
\label{sec:coda-comparision}

This article does not require a detailed understanding of Compositional Data
Analysis (CoDA). Still, CoDA has become increasingly popular for analyzing human
gut microbiome data, and as a result, we provide a brief comparison for readers
who are already familiar with that literature. A broader review of CoDA and a
more detailed introduction to CoDA in the context of microbiome data can be
found in \cite{greenacre2023codareview} and \cite{gloor2016allrelative},
respectively.

In brief, CoDA studies simplex-valued data: multivariate, continuous,
positive-valued vectors that sum to a constant value. In the context of
microbiome data, researchers often identify each datum \(\Y_{\cdot n}\) with a
single vector on simplex via normalization (e.g.,
\(\WC_{\cdot n}=\Y_{\cdot n}/\sum_{d} Y_{d n}\))~\citep{srinivasan2020knockoff}.
While we have used our notation \(\WC\), it is important to note that CoDA is
not formulated with any explicit notion of the system being measured. Said
differently, CoDA does not distinguish between the observed data and the
underlying system being studied. This is a critical difference between SRI and
CoDA. We use the system \(\W\) as a conceptual tool to define the true value of
our estimand \(\theta(\W)\). This, combined with our probabilistic definition of
compositional surveys, will allow us to investigate SRI and motivate our methods
using classical inferential criteria (e.g., consistency and type-I error
control). Lacking these structures, CoDA relies on three non-standard axioms to
motivate its approach~\citep[Section 2.2]{pawlowsky2015book}. Chief among these
is the principle of scale invariance, which states that any analysis of
compositional data should be invariant to the arbitrary scale of the data. As a
result of this axiom, CoDA restricts its attention to log-ratio estimands which
are scale invariant: e.g., estimands of the form
\(\gamma = f(\eta_{1}, \dots, \eta_{N})\) where
\(\eta_{n}=\log (\W_{in}/\W_{jn})\)~\citep{lovell2020counts}. In contrast, SRI
studies the estimation of scale reliant estimands.

Despite these critical differences, we choose terminology that is consistent
with CoDA. For example, we use the term \textit{scale} to refer to the sums
\(\WT_{n}=\sum_{d}\W_{dn}\) of the system in the same way that CoDA uses the
term to refer to the sums of the data \(\sum_{d}\Y_{dn}\). We choose the term
\textit{compositional survey} to distinguish our probability model from the
mathematical definition of compositional data while highlighting the intuitive
relationship between them. Note our definition of compositional surveys does not
require that the measurements \(\Y_{\cdot n}\) are continuous or sum to a
constant value. This is appealing as microbiome data lack information about the
system scale yet are count vectors that do not sum to a constant value.

The implications of these differences between CoDA and SRI are discussed in
later sections and summarized in Section~\ref{sec:discussion}.

\subsection{The Continuum between SII and SRI}
\label{sec:continuum}

Identifiability is not a dichotomous state but rather a continuum based on the
identification region of a parameter. This implies that there is a spectrum
between SRI and SII. Here, we define this spectrum before showing how these
definitions apply to inferential problems described in existing literature.

In the extreme case where \(\Id{\theta}P\) is a singleton set, \(\theta\) is
identified, and the estimation problem is \textit{SII}. At the opposite end
where \(\Id{\theta}P=\Theta\), \(\theta\) is \textit{fully} unidentified, and
the estimation problem is \textit{Full SRI}. Full SRI represents the extreme
case where, due to the arbitrary scale of the data, all values of
\(\theta\in \Theta\) are consistent with the observed measurements. In between
these two cases, where \(\Id{\theta}P\subsetneq \Theta\), \(\theta\) is
\textit{partially identified}, and the estimation problem is \textit{Partial
  SRI}.

In Section \ref{sec:limits-on-inference}, we show that the distinction between
SII, Full SRI, and Partial SRI is directly related to the conditions under which
certain types of inference (e.g., Frequentist versus Bayesian) are possible.
Before providing those results, we first motivate that all three types of
estimation problems are encountered within the literature. To do this, we
describe the two relevant factors that determine whether a problem is SII, Full
SRI, or Partial SRI. The factors are the degree to which the target estimand depends on
scale and the size of the identification region of \(\WT\) at
\(P_{\WC,\WT}(\Y)\).

\subsubsection{Characterization of Target Estimands}
\label{sec:estimands-classification}

Target estimands can be characterized based on the degree to which they depend on scale -- whether they are scale invariant, partially scale reliant, or fully scale reliant. We define this characterization before providing examples of each type. 

Let \(\WC \times \WTS\) denote the set of all systems that have a particular composition \(\WC\).
$\WTS$ is the space of scale vectors,  \(\mathbb{Z}^{N+}\) if \(W\) is integer valued or \(\mathbb{R}^{N+}\) if continuous. 
Let \(\theta(\WC \times \WTS\)) denote the image of this set with respect to the target estimand: 
\[\theta(\WC \times \WTS)=\{\theta\in \Theta: \exists\, \,  \WT\in \WTS, \theta(\WC, \WT)=\theta\}.\]
We define scale invariance/reliance based on the size of this image. If \(\theta(\WC \times \WTS)\) is a singleton set, we say that \(\theta\) is scale invariant at \(\WC\); otherwise, it is scale reliant at \(\WC\). We sub-categorize scale reliant estimands as fully scale reliant if \( \mbox{span}(\theta(\WC \times \WTS))=\Theta\) or partially scale reliant if \(\mbox{span}(\theta(\WC \times \WTS))\subsetneq \Theta\). Our definition reflects the following intuition. If \(\theta\) is scale invariant, then its value is uniquely determined by \(\WC\) alone and does not depend on \(\WT\). In contrast, if \(\theta\) is partially scale reliant then \(\WC\) limits the values that \(\theta\) can take on but does not uniquely determine its value. If \(\theta\) is fully scale reliant, then \(\WC\) places no limits on the values \(\theta\) can take on.

The scale reliance or scale invariance of a target estimand can depend on the composition \(\WC\). For example, in a companion article, we have studied a set of non-parametric estimands (based on modified Kolmogorov-Smirnov statistics) which is scale invariant for some \(\WC\) and scale reliant for others~\citep{mcgovern2023addressing}. For simplicity, this manuscript only considers target estimands where their scale reliance or scale invariance holds for all \(\WC\). 

\begin{example}[Microbiome Data Analyses]
The LFC estimand described in Equation~\eqref{eq:lfc} is fully scale reliant. This can be established as follows. For any \(\WC\) (and therefore any \(\thetaC_{d} \in (-\infty, \infty)\)) and any \(\theta_{d}\in (-\infty, \infty)\); there exists some \(\WT \in \WTS\), and corresponding \(\thetaT\in (-\infty, \infty)\), such that \(\theta_{d}=\thetaC+\thetaT\). That is, without restrictions on \(\thetaT\in (-\infty, \infty)\), fixing the value of \(\thetaC_{d}\) places no restrictions on the value of \(\theta_{d}\). 

At the opposite extreme, the types of log-ratio estimands purported by the field of compositional data analysis are scale invariant. For example, CoDA often analyzes log-ratio abundances between pairs of taxa  \(d_{i}\) and \(d_{j}\)~\citep{lovell2015proportionality}: 
\begin{equation}
  \label{eq:log-ratio-estimand}
  \eta_n = \log \frac{\W_{d_{i}n}}{\W_{d_{j}n}}.
\end{equation}
These quantities can be calculated from the system composition alone:
\begin{equation}
\eta_{n}=\log \frac{\W_{d_{i}n}}{\W_{d_{j}n}}=\log \frac{\WC_{d_{i}n}}{\WC_{d_{j}n}}.
\label{eq:ratio-decomposition}
\end{equation}

In between these two extremes there are partially scale reliant estimands. For example, in the analyse of human gut microbiome data a common task is to identify pairs of taxa whose abundances covary on the log-scale:
\begin{equation}
  \label{eq:cov-estimand}
  \Sigma_{d_{i}, d_{j}} = \text{cov} (\log \W_{d_{i}\cdot}, \log \W_{d_{j} \cdot}).
\end{equation}
These covariance estimands underlie the construction of ecological interaction networks~\citep{kurtz2015sparse,friedman2012sparcc}.
In Supplementary Section B.1, we prove this estimand is partially scale reliant: \(\WC\) provides a finite lower bound for this estimate but it is unbounded from above regardless of \(\WC\). 
\end{example}

\subsubsection{Characterization of Compositional Surveys}
\label{sec:survey-classification}

We categorize compositional surveys as partial or full compositional surveys depending on whether \(\WT\) is fully or partially identified by the model \(P_{\WC, \WT}(\Y)\). That is, partial compositional surveys provide some information about the system scale \(\Id{\WT} P\subsetneq \WTS\) whereas full compositional surveys provide no information \(\Id{\WT}P = \WTS\). For brevity, this article focuses on full compositional surveys. Still, we highlight this distinction as we believe that further study of partial compositional surveys would be beneficial.

\begin{example}
Authors who claim that microbiome data are compositional (e.g., \cite{gloor2017notoptional}) are essentially stating that the data provides so little information about \(\WT\) that can be equated with simplicial vectors. We interpret such claims as a belief that the data comes from a full compositional survey. Empirical studies support this view illustrating that there is little relationship between the system scale and the scale of the data~\citep{vandeputte2017quantitative,props2017absolute}. While we expect that future researchers will identify specific cases where microbiome data does provide some degree of information about scale; for simplicity, this article models microbiome data as full compositional surveys. 

Outside of microbiome studies there are clear cases where the data comes from partial compositional surveys. Consider political polling. Under standard conditions, a poll of 1000 voters in a town implies that the town must have at least 1000 voters. Still these data typically do not provide sufficient information to determine the true number of voters. 
\end{example}

\subsubsection{Theoretical Tools for Establishing SRI}
\label{sec:establishing-sri}
In Supplemental Section B.2 we provide theoretical results which establish how estimands and compositional surveys combine to determine if an estimation problem is Full SRI, Partial SRI, or SII. We use those results to draw the following conclusions about estimation from full compositional surveys (e.g., microbiome data). LFC estimation (Eq~\ref{eq:lfc}, e.g., differential abundance analysis) is Full SRI. Covariance estimation (Eq.~\ref{eq:cov-estimand}) is Partial SRI. Estimation of log-ratio quantities (e.g., Eq.~\ref{eq:log-ratio-estimand}) is SII. 

\subsection{Limits on Scale Reliant Inference}\label{sec:limits-on-inference}

Having formally defined SRI as a problem of identifiability, we can prove fundamental limits on the types of inference that can be performed using data with arbitrary scale. We frame these limits in terms of common Frequentist and Bayesian inferential criteria. We show that there are strict limits on Frequentist inference but none on Bayesian inference. However, rigorous Bayesian inference requires an uncommon type of model called a Bayesian Partially Identifiable Model (PIM). We use the concept of unacknowledged bias to discuss the practical implication of these limits. 

\subsubsection{Limits on Frequentist Inference} \label{sec:freq-limits}

We considers Frequentist inferential procedures motivated by canonical criteria such as consistency, unbiasedness, interval calibration, and type-I error control. The following four theorems follow from the definition of SRI as a problem of identifiability. Related impossibility proofs can be found
in other fields that encounter partially identified models: such as econometrics~\citep{dufour1997some,gabrielsen1978consistency} or in the study of error-in-variable models~\citep{gleser1987nonexistence}. Proofs are provided in Supplementary Section~B.3.

\begin{theorem}[Consistency]
  If \(\Id{\theta}P\) is not a singleton set (Full or Partial SRI), then there does not exist a sequence of estimators for the target estimand \(\theta\) that is consistent. 
\end{theorem}

\begin{theorem}[Unbiasedness]
  If \(\Id{\theta}P\) is not a singleton set (Full or Partial SRI), then there does not exist an estimator for the target estimand \(\theta\) that is unbiased. 
  \label{thm:bias}
\end{theorem}

\begin{theorem}[Calibration]
  Suppose there exists \(\theta\in \Theta\) and \(\lambda\in \Lambda\) such that \(Id_{\theta}(P)=\Theta\) (Full SRI). 
  There does not exist a non-trivial\footnote{See Supplementary Section B.3.} confidence set \(S(\Y)\) calibrated at level \(1-\alpha>0\).  
\end{theorem}

\begin{theorem}[Type-I Error Control]
  Suppose there exists some \(\theta\in \Theta\) and \(\lambda \in \Lambda\) such that \(Id_{\theta}(P)=\Theta\) (Full SRI). Then, for any \(\alpha>0\), there does not exist a non-trivial\footnote{See Supplementary Section B.3.} level-\(\alpha\) test of the null hypothesis \(\theta\in \Theta_{0}\subsetneq \Theta\). 
  \label{thm:type-i}
\end{theorem}


To the best of our knowledge, these are the first formal statements of what
types of inferential criteria can or cannot be attained when analyzing data with
arbitrary scale. These theorems provide formal justification for CoDA's
insistence on scale invariance, yet also show that existing arguments within the
CoDA community are overly restrictive.

Consistent with CoDA's arguments, these theorems show that the Frequentist ideal
is not attainable for Full SRI problems. For example, Theorem~\ref{thm:type-i},
contradicts a growing body of literature which claims to have developed rigorous
hypothesis tests for differential analysis of compositional data (Full SRI) that
control Type-I error~\citep{srinivasan2020knockoff, brill2022testing,
  love2014moderated, fernandes2014unifying}. In Supplementary Section C, we
review one paper that claims to have developed such a test and discuss how to
reconcile those claims with our results.

Still, our results show that CoDA's insistence on scale invariance goes to far.
For example, our results show that non-trivial hypothesis tests could be
developed for Partial SRI tasks. In fact, we have already done
so in \cite{mcgovern2023addressing}. Moreover, if we look beyond Frequentist
inference there are even more possibilities. In the next subsection, we show
that Bayesian inference is possible for all forms of SRI.

\subsection{Bayesian PIMs for SRI}
\label{sec:bayesian-limits}
\label{sec:unacknowledged-bias}

The previous section established strict limits on Frequentist inference in the
setting of SRI. That is, limits on inferential procedures motivated by common
criteria such as consistency. In contrast, this section explains why there are
no equivalent limits on Bayesian inference as standard Bayesian assumptions can
be satisfied. There are two principle assumptions which we review below: the
likelihood principle and the prior calibration principle. While we explain how
these assumptions can be satisfied, we also show that doing so requires a
specialized type of model called a Bayesian PIM.

Consider the problem of estimating LFCs for \(D\) taxa
(\(\theta=(\theta_{1}, \dots, \theta_{D})\)) from a microbiome study \(\Y\)
(Full SRI). To satisfy the prior calibration principle, the prior must
accurately represent prior beliefs about \(\theta\). In most real-world
scenarios we expect this would require a non-singular distribution
\(p(\theta)\), e.g., \(\text{LogNormal}(\theta \mid \mu, \Sigma)\). Using Equation \ref{eq:lfc-decomposition}, such priors can
reparameterized as \(p(\thetaC, \thetaT)\) where
\(\thetaC=(\thetaC_{1}, \dots, \thetaC_{D})\) is a vector-valued function of
\(\WC\) and \(\thetaT\) is a scalar function of \(\WT\). To satisfy the
likelihood principle, the chosen likelihood must accurately reflect the evidence
provided by the observed data \(\Y\) relevant to the model parameters
 \((\thetaC, \thetaT)\). However, since the observed
data comes from a full compositional survey, satisfying the likelihood principle requires a likelihood invariant to
\(\WT\):
$$\mathcal{L}(\WC, \WT ; \Y)\propto\mathcal{L}(\WC; \Y).$$ 
As an example, when analyzing multivariate count data, a product multinomial might be appropriate:
\[\mathcal{L}(\WC; \Y)=\prod_{n}\text{Multinomial}(\Y_{\cdot n} \mid  \WC_{\cdot n}).\]
For LFC estimation, since \(\thetaC\) is solely a function of \(\WC\) and
\(\thetaT\) is solely a function of \(\WT\), this could be reparameterized as a
likelihood of the form \(\mathcal{L}(\thetaC, \lambda; \Y)\) where \(\lambda\)
is some nuisance parameter that depends on \(\WC\). For brevity, we omit
\(\lambda\) from future notation. This combination of non-singular prior
\(p(\thetaC, \thetaT)\) with a likelihood that is invariant to one of the prior
parameters \(\mathcal{L}(\thetaC; \Y)\) is an example of a Bayesian PIM. A more
general introduction to Bayesian PIMs can be found in \cite{gustafson2015book}.

Bayesian PIMs can have posteriors that demonstrate appealing, yet
non-traditional behavior. Many traditional Bayesian models have posteriors which
concentrate to a point-mass asymptotically~\citep{schwartz1965bayes}. In SRI,
this behavior is undesirable. From Theorem~\ref{thm:bias}, any point
estimator is biased and therefore any Bayesian posterior that asymptotically
collapses to a point mass is problematic. Such methods are said to display an
\textit{unacknowledged bias} as they are biased yet fail to account for that
bias in their estimation of uncertainty~\citep[pg. 35]{gustafson2015book}. Unlike
those traditional methods, Bayesian PIMs often have non-zero asymptotic
variance. More concretely, consider that the posterior for the above model
factors as:
\begin{align}
p(\thetaC, \thetaT \mid  Y) &\propto \mathcal{L}(\thetaC ; Y) p(\thetaC, \thetaT)  \nonumber\\
&\propto \mathcal{L}(\thetaC; Y) p(\thetaC) p(\thetaT \mid  \thetaC) \nonumber\\
 &\propto p(\thetaC\mid  Y) p(\thetaT \mid  \thetaC). \label{eq:bayesian-pim-posterior}
\end{align}
So long as the prior is non-singular, the posterior factor
\(p(\thetaT \mid \thetaC)\) will not collapse, even asymptotically. The value of
a posterior that maintains non-zero variance is straight-forward: if the posterior is biased, then
only a posterior that has non-zero variance has any chance of covering the true
value. It is for this reason that Bayesian PIMs are often said to display an
\textit{acknowledged} rather than \textit{unacknowledged} bias. Reinforcing this
terminology, one author has even shown that the posterior of Bayesian PIM's can
be interpreted as a probabilistic model for the bias of the posterior
mean~\citep[pg. 35]{gustafson2015book}.

A major motivation of this work is to avoid unacknowledged bias. To underscore
the danger, consider that a hallmark of unacknowledged bias is a Type-I or
Type-II error rate that increases with more data (see Supplemental Figure 2 for
visual explanation). This behavior is counter to standard statistical wisdom --
where statistical procedures are supposed to improve with more data. For
concreteness, in Supplemental Section D we discuss three common pitfalls
encountered when developing Bayesian models for SRI including the use of Poisson
or Negative Binomial likelihoods. That section discusses how each pitfall can
lead to unacknowledged bias. In Section 4, we demonstrate the practical benefits
of Bayesian PIMs for microbiome data analysis using both simulated and real data
studies. Those studies demonstrate that common methods for microbiome
differential abundance analysis demonstrate unacknowledged bias. They also show
how Bayesian PIMs can drastically reduce Type-I and even Type-II error rates by
acknowledging their own bias.

\section{Scale Simulation Random Variables (SSRVs)}\label{sec:scale-simulation}

We now discuss practical solutions to SRI. We focus our attention on Bayesian PIM
approaches as they can be applied to both Partial and Full SRI tasks. Two challenges confront the use of Bayesian PIMs for SRI: model specification and computational tractability. 

Standard approaches to prior specification in Bayesian PIMs can lead to
unreasonable assumptions about the system scale. For LFC estimation, a prior of
the form \(p(\theta)\) may make implicit assumptions about \(\thetaT\) though
the conditional \(p(\thetaT\mid \thetaC)\). We demonstrate this pitfall in Section 4.2. 




Beyond model specification, Bayesian PIMs can be
challenging to infer. Returning to LFC estimation, note the posterior
factor \(p(\thetaT\mid \thetaC)\) retains the variance of the prior while the
variance of the factor \(p(\thetaC\mid \Y)\) will typically decrease with
increasing sample size. This leads to substantial anisotropy (i.e.,
\(\mathrm{var}(\thetaC\mid \Y) << \mathrm{var}(\thetaT\mid \Y)\)) which poses
computational challenges for traditional Bayesian inferential
tools~\citep{gustafson2015book}. Previous authors have suggested that tools
like importance sampling could address these problems~\citep{gustafson2015book}. In our
experience, those methods work poorly for SRI where posteriors are often
high-dimensional.

To address these challenges, we introduce Scale Simulation Random Variables
(SSRVs), a modeling framework that can help specify and infer Bayesian PIMs for
SRI.

\subsection{Defining SSRVs} \label{sec:define-SSRVs}

We start by defining SSRVs before discussing
how they relate to Bayesian PIMs and address the
aforementioned challenges. For simplicity, we limit our discussion to SSRVs for
full compositional surveys. Generalizations to partial compositional surveys are
straightforward yet may be more computationally challenging.

We define SSRVs as the combination of three components. The first is the target estimand, \(\theta(\WC, \WT)\). This estimand also defines the problem as SII versus SRI.  The second is the \textit{measurement model}, \(p(\WC\mid \Y)\). The measurement model represents posterior uncertainty in the system composition. 
The third component is a \textit{scale model}, \(p(\WT\mid \WC)\). The scale model is a prior distribution which represents uncertainty in the system scale and may be conditioned on the system composition.  With these components, a SSRV is a random variable \(\theta\) with probability distribution \(p(\theta\mid \Y)\) that can be defined via an integral transform.

 \begin{definition}[SSRVs: Integral Transform]
The following integral transformation defines a SSRV for \(\theta\) as the transformation of the scale and measurement models through the target estimand:
\begin{equation}
    P(\theta\in \set{A}\mid \Y)=\int_{(\WT, \WC)\in \theta^{-1}(\set{A})}p(\WT\mid \WC \, ) \, p(\WC\mid \Y) \, d\WC \, d\WT.
    \label{eq:integral-transform}
\end{equation}
 \end{definition}

The above integral transform suggests the following Monte-Carlo procedure to efficiently simulate SSRVs.  

\begin{algorithm}
  \SetAlgoLined\SetKwBlock{DoParallel}{do in parallel}{end}
  \KwData{\(\Y\)} \KwResult{\(S\) samples of \(\theta\)}
  \For{\(s\) in \(\{1, \dots, S\}\)}{
    Sample \(\WC[(s)] \sim p(\Y)\) \\
    Sample \(\WT[(s)] \sim p(\WC[(s)])\) \\
    Calculate \(\theta^{(s)}=\theta(\WC[(s)],\WT[(s)])\)
  }
  \caption{Monte-Carlo Algorithm for Simulating SSRVs}
  \label{alg:monte-carlo-sse}
\end{algorithm} 

 Less formally, SSRVs are simulations of the target estimand that incorporate both uncertainty in the system composition and uncertainty in the system scale. 
This interpretation is echoed in the SSRV mean which is the expectation of the target estimand taken with respect to the measurement and scale models:
\[\mathbb{E}[\theta\mid \Y]=\mathbb{E}_{p(\WT\mid \WC)}[\mathbb{E}_{p(\WC\mid \Y)}[\theta(\WC, \WT)]].\] 

\subsection{Relationship to Bayesian PIMs}
\label{sec:ssrv-pim-relationship}

SSRVs are Bayesian PIMs which are specified
using augmented variables \(\WT\) and \(\WC\). Within the integral of Equation \ref{eq:integral-transform}, the term
\(p(\WT\mid \WC)p(\WC\mid \Y)\) is the posterior of a
Bayesian PIM. Unlike the Bayesian PIMs discussed previously, the Bayesian PIM within the integral of Equation \ref{eq:integral-transform} uses \(\WT\) and \(\WC\) as parameters rather than \(\theta\). SSRVs
then use the target estimand to transform this posterior over \(\WC\) and \(\WT\) into a posterior over \(\theta\). SSRVs are simply an alternative method of specifying
Bayesian PIMs for SRI.

SSRVs mitigate the challenge of specifying and inferring Bayesian PIMs discussed
above. Whereas priors placed directly on \(p(\theta)\) may hide assumptions
about scale within the conditional \(p(\thetaT \mid \thetaC)\), the scale model
of an SSRV is that conditional. That is, assumptions about scale are an explicit
part of SSRV specification. In Section~\ref{sec:real-data} we highlight the
benefit of this transparency using real and simulated data examples. SSRVs can
also simplify posterior computations. Algorithm~\ref{alg:monte-carlo-sse} only
requires simulating from the measurement model and the scale model before
combining those simulations in the target estimand. The measurement model is a
standard (identified) Bayesian model, and the scale model is simply a prior.
Sampling from both distributions is comparatively straightforward and avoids the
anisotropy common with Bayesian PIMs.

\section{Applications}\label{sec:empirical-studies}

This section considers real and simulated case studies of microbiome differential abundance analysis to illustrate the theory and methods developed in the prior sections. We start by reviewing the ALDEx2~\citep{fernandes2014unifying} model which is a popular method for differential abundance analysis. We show that ALDEx2 can be thought of as an SSRV with an unrealistic scale model that underestimates prior uncertainty. Using real and simulated studies we show how this scale model leads to elevated rates of type-I and type-II errors in practice. Remarkably, we find similar patterns of false positives and false negatives with the distinct DESeq2 model~\citep{love2014moderated} suggesting our insights generalize beyond ALDEx2. We show how SSRVs address the pitfalls of these methods.

\subsection{ALDEx2 is a SSRV with a Degenerate Scale Model}
\label{aldex2-as-an-ssrv}

While there are many popular tools for differential abundance analysis, we
highlight ALDEx2~\citep{fernandes2014unifying} as it has a particularly simple
mathematical form which aligns closely with SSRVs.

In Supplementary Section D.3, we present the ALDEx2 model as originally published
in \cite{fernandes2014unifying}. For brevity and clarity, we
briefly review a simplified form of the ALDEx2 model which can
be described in three steps. In the first step, posterior draws
from independent multinomial-Dirichlet models are used to simulate uncertainty
in the composition conditioned on the observed data. In the second step, those
draws are normalized using the Centered Log-Ratio (CLR) transform which is a
  transform from the \(D\)-simplex (\(\mathbb{S}^{D}\)) to \(\mathbb{R}^{D}\)
  defined as
  \(\phi(x)=\log x - \mathbf{1}\mean(\log x)\)  for a vector \(x\in \mathbb{S}^{D}\). In the third step, LFCs are calculated using the CLR-normalized
relative abundances. The result is a distribution over LFCs. In fact, these
three steps correspond to the measurement model, scale model, and target
estimand of a SSRV.

The multinomial-Dirichlet samples imply a measurement model of the form:
\[\WC_{\cdot n} \sim \text{Dirichlet}(\Y_{\cdot n}+\alpha)\]
where \(\alpha\) are the chosen Dirichlet prior parameters (default to \(\alpha=0.5\cdot\mathbf{1}_{D}\) in ALDEx2). The target estimand is the LFC estimand (Eq. \ref{eq:lfc}):
\[\theta_{d}=\mean_{n:x_{n}=1}\log \W_{dn}-\mean_{n:x_{n}=0}\log \W_{dn}.\] 
Together, these two components suggest that the CLR normalization step represents some form of implicit scale model. 
This normalization can be written as:
\[\log \W_{dn}=\log \WC_{dn}-\mean\left(\log \WC_{\cdot n}\right).\]
Comparing this to \(\log \W_{dn}=\log \WC_{dn} + \log \WT_{n}\) implied by Equation \ref{eq:lfc-decomposition}, it is clear that this normalization equates to a scale model of the form:
\[\log \WT_{n} \sim \delta[-\mean \left(\log \WC_{\cdot n}\right)]\]
where \(\delta[x]\) denotes a point mass centered at the point \(x\). This scale model has zero variance. That is, ALDEx2
does not account for any uncertainty in its implicit assumption regarding the
unmeasured system scale. Errors in this assumption lead to biased estimates. 

As an aside, we expect that similar unacknowledged bias due to errors in
implicit assumptions regarding scale may be found in other fields.
Normalizations are common when analyzing polling data~\citep{topirceanu2020polling} and in
Supplementary Section F, we provide an example of an implicit assumption in a
specific type of ecological survey.

\subsection{ALDEx2's Degenerate Scale Model Drives False Positives}
\label{sec:relaxing-aldex2}

We simulated a microbiome study comparing human gut microbiota in otherwise healthy individuals before and after treatment with a mild, narrow spectrum antibiotic whose true effect is to slightly decrease the abundance of 4 of the 20 simulated taxa (full simulation details provided in Supplementary Section G.1). Figure \ref{fig:aldex_ssrv}a shows the simulated true abundances (\(\W\)) for a set of \(N=100\) study participants equally split between the pre- and post-antibiotic conditions. We simulated data \(\Y\) with arbitrary scale by resampling the true abundances \(\W\) with a multinomial distribution so the arbitrary scale of each sequenced sample \(\Y_{\cdot n}\) has a total of \(5,000\) total counts. We applied two popular DA models (DESeq2~\citep{love2014moderated} and ALDEx2~\citep{fernandes2014unifying}) to \(\Y\) and, in Figure \ref{fig:aldex_ssrv}, show the pattern of true/false positives/negatives that result. In terms of Type-I Error control, both methods are worse than determining statistical significance by flipping a fair coin. Moreover, in Supplementary Figure~3 we show that the false discovery rate of both methods increases with increasing sample size -- the hallmark of unacknowledged bias. 

\begin{figure}
    \centering
    \includegraphics[width=5.6in]{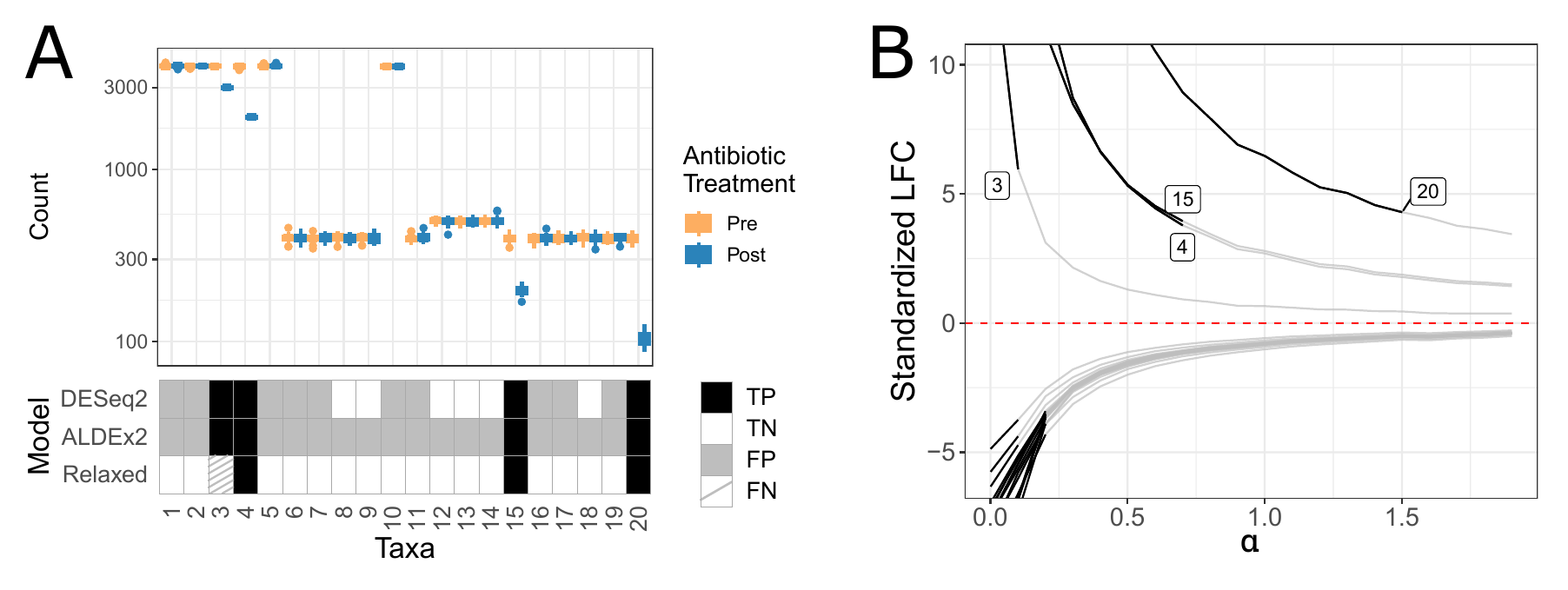}
    \caption{\footnotesize (a) Modeling results for DESeq2, ALDEx2, and a SSRV model.
      The \textit{Relaxed} scale model is centered on the same geometric assumption as CLR normalization assumption but considers error in that assumption by using $\alpha = 0.5$.
      b) Sensitivity analysis showing the estimated standardized log-fold change (LFC) for each taxon under the Relaxed model as a function of $\alpha$. The average and standard deviation of log fold change was computed across Monte Carlo samples, and a standardized value was computed for each entity by dividing the mean LFC by its corresponding standard deviation. A black line signifies that the estimated effect size was statistically different from zero at 95\% confidence. The four taxa that were truly differentially expressed are labeled (Taxa 3, 4, 15, and~20).}
    \label{fig:aldex_ssrv}
\end{figure}

As described in the prior section, we expect that the implicit scale model used
by ALDEx2 underestimates a researcher's true uncertainty in the system scale
leading to an unacknowledged bias and driving the false positives seen in Figure
\ref{fig:aldex_ssrv}. To evaluate this hypothesis, we created a relaxation of the
ALDEx2 model replacing the degenerate (point-mass) scale model with the
following relaxation:
\(\log \WT_{n}\sim N(-\mean \left(\log \WC_{\cdot n}\right), \alpha^{2})\). 

Our relaxation is sufficient to remove all of ALDEx2's false positives and
retain most if not all of the true positives (Figure~\ref{fig:aldex_ssrv}a).
Additionally, our relaxation addresses the unacknowledged bias and does not lead
to the same asymptotic increase in false discovery rates as seen with ALDEx2 and
DESeq2 (Supplemental Figure~3). Moreover, sensitivity analysis shows that our
results are largely insensitive to the choice of \(\alpha\): even small amounts
of uncertainty are sufficient to remove the majority of false positives at the
cost of just a single false negative (taxon 3; Figure~\ref{fig:aldex_ssrv}b).
Only when \(\alpha\) is chosen to be implausibly large (\(\alpha>1.4\)) does the
relaxed ALDEx2 model fail to find true positives. Considering that the CLR
estimate of scale corresponds to \(\hat{\theta}^{\perp}_{CLR}=0.03\) a value of
\(\alpha>1.4\) reflects a greater than 5\% probability that the log-fold-change
of scales, \(\theta^{\perp}\), is outside of the range
\([\exp^{-1.96\times 1.4^2 + \hat{\theta}^{\perp}_{CLR}}, \exp^{1.96\times 1.4^2 +
  \hat{\theta}^{\perp}_{CLR}}]=[0.02, 48.02]\); which is implausible for a
narrow spectrum antibiotic. In summary, our results indicate that the CLR
normalization in ALDEx2 drives its observed false positives and that these can
be mitigated by considering potential errors in assumptions about scale. The
observed pattern of false positives within DESeq2 suggests similar effects may
be at play within that model even though it uses a different median-of-ratios
normalization.

\subsection{Application to a novel 16S rRNA-Seq dataset}\label{sec:real-data}

Although artificial gut systems (bioreactors) are seeded with human feces and often used as a model for
host-associated microbiota, many bioreactors undergo a two week equilabration period in which
the inoculated communities diverge from those seen in human hosts and adapt to their new \textit{ex
  vivo} environment. To evaluate the limits of these model systems, we ran and collected data from 8
bioreactors over the course of the first two weeks after inoculation of which 6 were used for analyses due to experimental constraints (see Supplemental Section G.2). Each bioreactor was seeded
with 100mL of fecal slurry and maintained at a constant 400mL of growth media over the course of the
experiment. Our primary interest was in determining which of the microbial taxa present in the
innoculum were able to grow in the bioreactors. We quantify this interest using the differential
abundance target estimator given in Equation~\eqref{eq:lfc} with \( x_{n}=0 \) denoting day 1 of the
experiment and \( x_{n}=1 \) denoting day 14. In this case, we were primarily interested in the
\textit{de facto} total scale, i.e., we wanted to identify taxa that increased or decreased in total
cellular number over the two week period.

This data is ideal for our study as our experimental system allowed us to obtain
high-quality measurements of scale in this bioreactors. Because each vessel was
homogeneously mixed, microbial concentration measured from a sample of the
vessel via flow cytometry was presumably identical to microbial concentration
within the vessel. Moreover, because each vessel was maintained at a constant
400mL volume throughout the course of the experiment, changes in microbial
concentration between vessels or between time-points were proportional to
changes in total microbial load. Flow cytometry measurements were obtained to
measure total cellular concentration. These measurements were taken in
triplicate for each 16S rRNA sequencing measurement to allow us to quantify
measurement noise. Further experimental details are presented in Supplemental
Section G.2.

Five models were compared on the resultant data: one specified as a Bayesian PIM, two specified as SSRVs,
ALDEx2, and DESeq2. We used the following multinomial log-normal Bayesian PIM,
which is an extension of the Bayesian multinomial logistic-normal models that
have become popular for analysis of microbiota survey data
\citep{silverman2022bayesian}:
\begin{align}
  \Y_{\cdot n} &\sim \text{Multinomial}(\WC_{\cdot n}) \\
  \W_{\cdot n} &= \exp({\psi_{\cdot n}}) \\
  \psi_{\cdot n} &\sim N(\mathbf{B} X_{\cdot n},\boldsymbol{\Omega})
\end{align}
with priors
\(\mathbf{B} \sim N(\mathbf{M},\boldsymbol{\Omega}, \boldsymbol{\Gamma})\) and
\(\boldsymbol{\Omega} \sim IW(u, \boldsymbol{\Xi})\). This is a Bayesian PIM
since the multinomial likelihood is only parameterized in terms of \( \WC \)
whereas prior is defined over both \(\WC\) and \(\WT\). We chose the prior
hyper-parameters to echo standard defaults we have observed in the literature,
i.e., we chose \(\mathbf{M}=0\) and choose \(\mathbf{\Gamma}\) and \(\mathbf{\Xi}\) to be diagonal
matrices (see Supplementary Section G.2 for full prior specifications).

There are two critical limitations of this Bayesian PIM. First, it is computationally
infeasible to infer this model directly. Instead, to obtain posterior samples of
the parameter \(\mathbf{B}\), we had to represent this model as an SSRV and sample
\(\mathbf{B}\) using Algorithm~\ref{alg:monte-carlo-sse}. Supplementary
Section~H derives the SSRV representation of this Bayesian PIM and shows that
the measurement model is a Bayesian multinomial logistic-normal model and the
scale model is a log-normal linear model. The second, more important, limitation
of this model is that the assumptions about scale are not transparent. In fact,
this seemingly innocuous and standard prior specification actually implies
strong, biologically implausible assumptions: implicit in this prior is a
non-trivial conditional relationship of the form \(p(\thetaT\mid \thetaC)\).
This leads to a posterior for \(\thetaT\) which reflects a strong, biologically
unreasonable belief that total microbial load is higher on day 1 than on day 14
(see Supplementary Section G.2). Below we show this leads to unacknowledged bias
which presents as elevated Type-I and Type-II error rates.

To illustrate how SSRVs can enhance model specification and ease prior
specification, we designed two SSRVs using the same measurement model as the
above Bayesian PIM but with two different scale models. Letting \( n\in\{1, \dots, N\}\) denote a
given vessel at a given time-point, these two scale models were both log-normal models of the form :
\begin{equation}
\log \WT_{n} \sim N(c + \mu_{n}, \alpha_{n}^{2})
\end{equation}
where \(c\) is a constant and \( \mu \) is subject to the constraint
\(\sum_{n=1}^{N}\mu_{n}=0\). Note that the LFC target estimand given by Equation
\ref{eq:lfc} is invariant to the value of \(c\) and so we can set \(c=0\) with
no loss of generality.

We created the \textit{Design} scale model based only on prior knowledge of the experimental design. 
Since we know that each vessel contained 400mL of growth
media and the innoculum consisted of 100 mL of fecal slurry, we expect that the
total microbial load would, on average, increase from the day 1 to day 14 by a
factor of approximately 4. We encapsulated these beliefs in the scale model
\[\frac{f(\WT_{{j}, \text{t=1}})}{f(\WT_{{j}, \text{t=14}})} \sim
  \mathrm{LogNormal}(\bar{d}, \tau^2) \] where $\bar{d} = \log(100/400)$ and
$\tau = 0.5$.

Finally, we created the \textit{Flow} scale model built from the flow-cytometry
measurements. In this model \( \mu_{n} \) and \( \alpha^{2}_{n} \) were
estimated as mean and variance our flow cytometry measurements on the log scale.
We consider the SSRV created using this scale model our gold standard against
which we define false positives. For concreteness, we define a result as a false positive in two cases: first, that the result is significant in a given model but not in the gold standard and second, that the result is significant in both the model and the gold standard, but the two models disagree on sign.

\begin{figure}[t]
    \centering
	\includegraphics[width=4.5in]{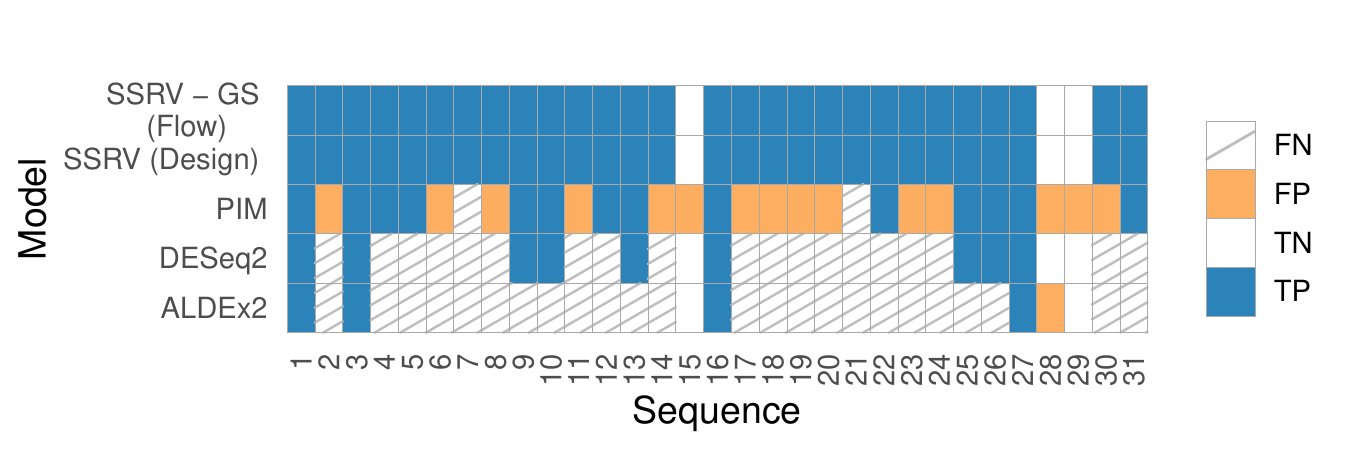}
	\caption{\footnotesize Comparison of results from ALDEx2, DESeq2, the Bayesian PIM, and two SSRVs using
          real data from an \textit{ex vivo} artificial gut study. ``GS''
          denotes the gold-standard model against which true/false
          positives/negatives are defined. Our gold standard (the SSRV using the
          Flow scale model) is the only model that could use the replicate
          flow-cytometry measurements collected in this study. For brevity, taxa
          that were identified as differentially abundant with respect to any of
          the 5 models are not shown (based on p-value less than 0.05 or
          posterior 95\% credible interval not covering zero).}
	\label{fig:ssrv-compare}
    \end{figure}

We applied the Bayesian PIM, the two SSRVs, ALDEx2, and DESeq2 to the artificial
gut data; the results are summarized in Figure~\ref{fig:ssrv-compare}. As before
we find that ALDEx2 and DESeq2 have high inferential error rates. However,
unlike the prior study where these were primarily Type-I errors, here we observe
far more Type-II errors owing to the fact that most taxa are actually increasing
in abundance as the community grows to occupy the larger vessels and this study
has a relatively small sample size (\(N=12\)). Still we find the hallmark of
unacknowledged bias: using resampling, Supplemental Figure 5 shows that ALDEx2
and DESeq2 display increasing rates of false discoveries with increasing sample
size while consistently displaying Type-II error rates far in excess of the
Design model. For example with just 20 vessels, the FDR of ALDEx2 and DESeq2 are
approximately 30\% and 15\% respectively. In contrast, even at 50 vessels, the
Design model controls FDR at a nominal 5\%. As expected based on our
investigation of the Bayesian PIM prior, the Bayesian PIM model displays a false discovery
rate of 45\% which further increases with more data (Supplemental Figure 5).

\begin{figure}[t]
	\centering
	\includegraphics[width = 3.5in]{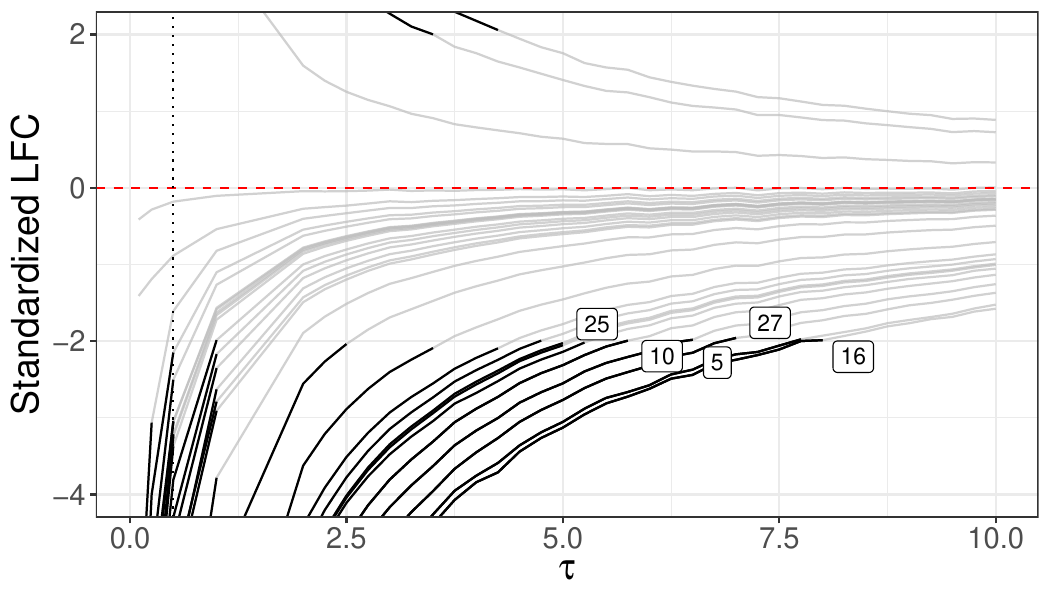}
	\caption{\footnotesize Sensitivity analysis of the Design SSRV as a function of the
          parameter $\tau$ in the Design scale model. The mean, standard
          deviation, and 95\% credible interval were computed using the
          posterior distribution of log-fold change for each entity. The
          standardized log-fold change (LFC) was calculated by dividing the mean
          log-fold change by the corresponding estimate for standard deviation.
          The standardized LFC as a function of \(\tau\) is shown for each taxon
          in the artificial gut study and is colored black if the corresponding
          credible interval excluded zero and grey otherwise. As an
          illustration, we highlight the 5 taxa that are the most clearly
          differentially abundant in the sense that their credible sets exclude
          zero over the largest range of \(\tau\). For comparison, we depict the
          value of \(\tau\) used in Figure \ref{fig:ssrv-compare} (\(\tau=0.5\))
          as a black dotted line.}
	\label{fig:tau-design-ssrv}
\end{figure}

Finally, Figure~\ref{fig:tau-design-ssrv} shows that the major conclusions of
this analysis, namely the remarkably performance improvement of the Design model
compared to alternatives is largely insensitive to the choice of the parameter
\(\tau\). So long as $\tau$ is greater than 0.35, the FDR of the Design model will be less
than 3.5\%. Moreover, so long as we choose $\tau$ less than 5, the Type-II error
rate of the design-based SSRV will be less than that of either DESeq2 or ALDEx2.
A value of \(\tau\geq 5\) would require we accept, with 2.5\% probability, that
the microbial load at day 14 was at least
\(4\exp(1.96*5^{2})\approx 7.6 \times 10^{21}\) times greater than at day 1.
Even if there was only one microbe at Day 1, this is not physically possible: as
bacterium weights approximately \(10^{-12}\) grams, this would require the
vessels at day 14 to weight more than~\(10^{6}\)~kilograms.

\section{Discussion}\label{sec:discussion}

In this article, we have framed the challenge of data that lacks scale
information as an estimation problem involving partially identified models.
Through this formulation, we have been able to provide the first formal limits
on inference with these data. Our results are consistent with the CoDA
perspective: they show that care is needed as spurious scientific conclusions
can be reached due to the lack of information about the system scale contained
in these data. However, unlike the CoDA perspective, our results argue that a
dogmatic restriction to scale invariance is overly prohibitive. Consider that,
even if data is compositional (i.e., a full compositional survey with simplex
valued measurements), Bayesian PIMs can be used to perform SRI while satisfying
Bayesian ideals. In fact, hypothesis tests and confidence intervals which
satisfy Frequentist ideals can be constructed from compositional data if the
estimand is partially scale reliant. Overall, we argue that SRI is possible yet
requires new types of statistical methods that can account for the limited
information contained in the observed data.

Beyond discussing limits on inference, this article introduces the class of
Bayesian PIMs for SRI. There is an interesting connection between the SSRV
formulation of Bayesian PIMs and adversarial robustness. Adversarial methods are
well known in the machine learning literature as a way of enhancing the
robustness or \textit{stability}~\citep{yu2020veridical} of predictions. For
example, within computer vision, noise is sometimes added to images in order to
ensure that trained models are stable and are not sensitive to slight variation
when applied out-of-sample \citep{santos2022avoiding}. Due to concerns about
rigor and reproducibility in science, some authors have advocated that
adversarial perturbations can be applied to ensure robustness of statistical
inference \citep{yu2020veridical}. Within SSRVs, the use of the scale model can
be viewed as a type of adversarial perturbation: for each sample from the SSRV
we apply a different random perturbation sampled from the scale model. One way
of understanding the utility of Bayesian PIMs (and SSRVs) is that they involve a
form of adversarial robustness to perturbations of scale.

In this article, we have focused on Bayesian approaches to SRI as these methods,
unlike Frequentist methods, remain valid even in the setting of Full SRI. To
clarify the problem with Frequentist methods, consider the problem of confidence
estimation. Intuitively, a level-\(\alpha\) confidence set for an estimand
\(\theta\) should, with probability \(1-\alpha\), cover the identification region
\(\Id{\theta} P\). The problem is that, for Full SRI, \(\Id{\theta} P=\Theta\)
and therefore any level-\(\alpha\) confidence set must be uselessly large. Yet,
for Partial SRI, useful confidence estimation (and hypothesis testing) may be
possible as \(\Id {\theta} P\subsetneq \Theta\). For example, we recently proved
that useful and robust hypothesis testing is possible for a particular partial
SRI problem called differential set analysis~\citep{mcgovern2023addressing}.
Overall, we believe that Frequentist SRI for Partial SRI problems is an
intriguing and likely fruitful area for future research. In fact, there is
already a rich literature on Frequentist confidence estimation for partially
identified models can likely be adapted to the task. We refer interested readers
to \cite{imbens2004confidence} and \cite{stoye2009more} for a thorough review of
those methods.

For simplicity, we have focused on compositional surveys which are a subset of
data with arbitrary scale where the system composition is identifiable. Yet in
many practical applications, we expect that the system composition may not be
identifiable as the measurement process may involve unobserved compositional
biases. For example, in the 16S rRNA microbiome surveys, the PCR amplification
process can lead to an over-representation of some taxa compared to others in
the sequenced pool~\citep{silverman2021measuring}. If compositional bias is substantial, then it may
tighten the limits on inference further restricting the types of inference that
can be performed rigorously with these data. Just as Bayesian PIMs can be used
to address arbitrary scale, we expect these same models may be adapted to
provide robustness to compositional bias. Overall, we believe that the study of
compositional bias represents a fruitful area for future research.

\begin{acks}[Data Access and Availability]
All data and \texttt{R} scripts necessary to replicate all analyses in the manuscript and supplement are located at \url{https://github.com/michellepistner/ScaleSimulation}.
\end{acks}

\begin{acks}[Acknowledgments]
We thank Rachel Silverman, Juan Jose Egozcue, Vera Pawlosky-Glahn, Gregory
Gloor, Michael Love, and James Morton for their manuscript comments.
\end{acks}

\begin{funding}
  JDS and MPN were supported in part by NIH 1R01GM148972-01. LAD acknowledges
  support from the the Global Probiotics Council, a Searle Scholars Award, the
  Hartwell Foundation, an Alfred P. Sloan Research Fellowship, the Translational
  Research Institute through Cooperative Agreement NNX16AO69A, the Damon Runyon
  Cancer Research Foundation, and NIH 1R01DK116187-01. SM would like to
  acknowledge the support of grants NSF IIS-1546331, NSF DMS-1418261, NSF
  IIS-1320357, NSF DMS-1045153, and NSF DMS-1613261.
\end{funding}

\begin{supplement}
\stitle{Supplement to ``Scale Reliant Inference''}
\sdescription{Includes all proofs, supplementary figures, discussions, and experimental details described in the main text.}
\end{supplement}


\bibliographystyle{imsart-nameyear} 
\bibliography{refs}       

\begin{thebibliography}{40}

\bibitem[\protect\citeauthoryear{Billheimer, Guttorp and
  Fagan}{2001}]{billheimer2001statistical}
\begin{barticle}[author]
\bauthor{\bsnm{Billheimer},~\bfnm{Dean}\binits{D.}},
  \bauthor{\bsnm{Guttorp},~\bfnm{Peter}\binits{P.}} \AND
  \bauthor{\bsnm{Fagan},~\bfnm{William~F}\binits{W.~F.}}
(\byear{2001}).
\btitle{Statistical interpretation of species composition}.
\bjournal{Journal of the American Statistical Association}
\bvolume{96}
\bpages{1205--1214}.
\end{barticle}
\endbibitem

\bibitem[\protect\citeauthoryear{Brill, Amir and
  Heller}{2022}]{brill2022testing}
\begin{barticle}[author]
\bauthor{\bsnm{Brill},~\bfnm{Barak}\binits{B.}},
  \bauthor{\bsnm{Amir},~\bfnm{Amnon}\binits{A.}} \AND
  \bauthor{\bsnm{Heller},~\bfnm{Ruth}\binits{R.}}
(\byear{2022}).
\btitle{Testing for differential abundance in compositional counts data, with
  application to microbiome studies}.
\bjournal{The Annals of Applied Statistics}
\bvolume{16}
\bpages{2648--2671}.
\end{barticle}
\endbibitem

\bibitem[\protect\citeauthoryear{Caporaso et~al.}{2011}]{caporaso2011global}
\begin{barticle}[author]
\bauthor{\bsnm{Caporaso},~\bfnm{J.~Gregory}\binits{J.~G.}},
  \bauthor{\bsnm{Lauber},~\bfnm{Christian~L.}\binits{C.~L.}},
  \bauthor{\bsnm{Walters},~\bfnm{William~A.}\binits{W.~A.}},
  \bauthor{\bsnm{Berg-Lyons},~\bfnm{Donna}\binits{D.}},
  \bauthor{\bsnm{Lozupone},~\bfnm{Catherine~A.}\binits{C.~A.}},
  \bauthor{\bsnm{Turnbaugh},~\bfnm{Peter~J.}\binits{P.~J.}},
  \bauthor{\bsnm{Fierer},~\bfnm{Noah}\binits{N.}} \AND
  \bauthor{\bsnm{Knight},~\bfnm{Rob}\binits{R.}}
(\byear{2011}).
\btitle{Global patterns of {16S} {rRNA} diversity at a depth of millions of
  sequences per sample}.
\bjournal{Proceedings of the National Academy of Sciences}
\bvolume{108}
\bpages{4516--4522}.
\bdoi{10.1073/pnas.1000080107}
\end{barticle}
\endbibitem

\bibitem[\protect\citeauthoryear{Dos~Santos
  et~al.}{2024}]{dossantos2024aldexvaginal}
\begin{barticle}[author]
\bauthor{\bsnm{Dos~Santos},~\bfnm{Scott~J.}\binits{S.~J.}},
  \bauthor{\bsnm{Copeland},~\bfnm{Clara}\binits{C.}},
  \bauthor{\bsnm{Macklaim},~\bfnm{Jean~M.}\binits{J.~M.}},
  \bauthor{\bsnm{Reid},~\bfnm{Gregor}\binits{G.}} \AND
  \bauthor{\bsnm{Gloor},~\bfnm{Gregory~B.}\binits{G.~B.}}
(\byear{2024}).
\btitle{Vaginal metatranscriptome meta-analysis reveals functional BV subgroups
  and novel colonisation strategies}.
\bjournal{bioRxiv}.
\bdoi{10.1101/2024.04.24.590967}
\end{barticle}
\endbibitem

\bibitem[\protect\citeauthoryear{Dufour}{1997}]{dufour1997some}
\begin{barticle}[author]
\bauthor{\bsnm{Dufour},~\bfnm{Jean-Marie}\binits{J.-M.}}
(\byear{1997}).
\btitle{Some impossibility theorems in econometrics with applications to
  structural and dynamic models}.
\bjournal{Econometrica: Journal of the Econometric Society}
\bpages{1365--1387}.
\end{barticle}
\endbibitem

\bibitem[\protect\citeauthoryear{Fernandes
  et~al.}{2014}]{fernandes2014unifying}
\begin{barticle}[author]
\bauthor{\bsnm{Fernandes},~\bfnm{Andrew~D}\binits{A.~D.}},
  \bauthor{\bsnm{Reid},~\bfnm{Jennifer~Ns}\binits{J.~N.}},
  \bauthor{\bsnm{Macklaim},~\bfnm{Jean~M}\binits{J.~M.}},
  \bauthor{\bsnm{McMurrough},~\bfnm{Thomas~A}\binits{T.~A.}},
  \bauthor{\bsnm{Edgell},~\bfnm{David~R}\binits{D.~R.}} \AND
  \bauthor{\bsnm{Gloor},~\bfnm{Gregory~B}\binits{G.~B.}}
(\byear{2014}).
\btitle{Unifying the analysis of high-throughput sequencing datasets:
  characterizing {RNA-seq}, {16S} {rRNA} gene sequencing and selective growth
  experiments by compositional data analysis}.
\bjournal{Microbiome}
\bvolume{2}
\bpages{1--13}.
\end{barticle}
\endbibitem

\bibitem[\protect\citeauthoryear{Friedman and Alm}{2012}]{friedman2012sparcc}
\begin{barticle}[author]
\bauthor{\bsnm{Friedman},~\bfnm{Jonathan}\binits{J.}} \AND
  \bauthor{\bsnm{Alm},~\bfnm{Eric~J.}\binits{E.~J.}}
(\byear{2012}).
\btitle{{Inferring Correlation Networks from Genomic Survey Data}}.
\bjournal{PLoS Computational Biology}
\bvolume{8}
\bpages{e1002687}.
\bdoi{10.1371/journal.pcbi.1002687}
\end{barticle}
\endbibitem

\bibitem[\protect\citeauthoryear{Gabrielsen}{1978}]{gabrielsen1978consistency}
\begin{barticle}[author]
\bauthor{\bsnm{Gabrielsen},~\bfnm{Arne}\binits{A.}}
(\byear{1978}).
\btitle{Consistency and identifiability}.
\bjournal{Journal of Econometrics}
\bvolume{8}
\bpages{261--263}.
\end{barticle}
\endbibitem

\bibitem[\protect\citeauthoryear{Gleser and
  Hwang}{1987}]{gleser1987nonexistence}
\begin{barticle}[author]
\bauthor{\bsnm{Gleser},~\bfnm{Leon~Jay}\binits{L.~J.}} \AND
  \bauthor{\bsnm{Hwang},~\bfnm{Jiunn~T}\binits{J.~T.}}
(\byear{1987}).
\btitle{The nonexistence of 100 (1-$\alpha$)\% confidence sets of finite
  expected diameter in errors-in-variables and related models}.
\bjournal{The Annals of Statistics}
\bpages{1351--1362}.
\end{barticle}
\endbibitem

\bibitem[\protect\citeauthoryear{Gloor et~al.}{2016a}]{gloor2016allrelative}
\begin{barticle}[author]
\bauthor{\bsnm{Gloor},~\bfnm{Gregory~B}\binits{G.~B.}},
  \bauthor{\bsnm{Wu},~\bfnm{Jia~Rong}\binits{J.~R.}},
  \bauthor{\bsnm{Pawlowsky-Glahn},~\bfnm{Vera}\binits{V.}} \AND
  \bauthor{\bsnm{Egozcue},~\bfnm{Juan~Jos{\'e}}\binits{J.~J.}}
(\byear{2016}a).
\btitle{It's all relative: analyzing microbiome data as compositions}.
\bjournal{Annals of {E}pidemiology}
\bvolume{26}
\bpages{322--329}.
\end{barticle}
\endbibitem

\bibitem[\protect\citeauthoryear{Gloor et~al.}{2016b}]{gloor2016uncertainty}
\begin{barticle}[author]
\bauthor{\bsnm{Gloor},~\bfnm{Gregory~Brian}\binits{G.~B.}},
  \bauthor{\bsnm{Macklaim},~\bfnm{Jean~M}\binits{J.~M.}},
  \bauthor{\bsnm{Vu},~\bfnm{Michael}\binits{M.}} \AND
  \bauthor{\bsnm{Fernandes},~\bfnm{Andrew~D}\binits{A.~D.}}
(\byear{2016}b).
\btitle{Compositional uncertainty should not be ignored in high-throughput
  sequencing data analysis}.
\bjournal{Austrian Journal of Statistics}
\bvolume{45}
\bpages{73--87}.
\end{barticle}
\endbibitem

\bibitem[\protect\citeauthoryear{Gloor et~al.}{2017}]{gloor2017notoptional}
\begin{barticle}[author]
\bauthor{\bsnm{Gloor},~\bfnm{Gregory~B}\binits{G.~B.}},
  \bauthor{\bsnm{Macklaim},~\bfnm{Jean~M}\binits{J.~M.}},
  \bauthor{\bsnm{Pawlowsky-Glahn},~\bfnm{Vera}\binits{V.}} \AND
  \bauthor{\bsnm{Egozcue},~\bfnm{Juan~J}\binits{J.~J.}}
(\byear{2017}).
\btitle{Microbiome datasets are compositional: and this is not optional}.
\bjournal{Frontiers in Microbiology}
\bvolume{8}
\bpages{2224}.
\end{barticle}
\endbibitem

\bibitem[\protect\citeauthoryear{Greenacre
  et~al.}{2023}]{greenacre2023codareview}
\begin{barticle}[author]
\bauthor{\bsnm{Greenacre},~\bfnm{Michael}\binits{M.}},
  \bauthor{\bsnm{Grunsky},~\bfnm{Eric}\binits{E.}},
  \bauthor{\bsnm{Bacon-Shone},~\bfnm{John}\binits{J.}},
  \bauthor{\bsnm{Erb},~\bfnm{Ionas}\binits{I.}} \AND
  \bauthor{\bsnm{Quinn},~\bfnm{Thomas}\binits{T.}}
(\byear{2023}).
\btitle{Aitchison's {Compositional} {Data} {Analysis} 40 {Years} {On}: {A}
  {Reappraisal}}.
\bnote{arXiv:2201.05197 [stat]}.
\end{barticle}
\endbibitem

\bibitem[\protect\citeauthoryear{Gustafson}{2015}]{gustafson2015book}
\begin{bbook}[author]
\bauthor{\bsnm{Gustafson},~\bfnm{Paul}\binits{P.}}
(\byear{2015}).
\btitle{{Bayesian inference for partially identified models: Exploring the
  limits of limited data}}
\bvolume{140}.
\bpublisher{CRC Press}.
\end{bbook}
\endbibitem

\bibitem[\protect\citeauthoryear{Imbens and
  Manski}{2004}]{imbens2004confidence}
\begin{barticle}[author]
\bauthor{\bsnm{Imbens},~\bfnm{Guido~W}\binits{G.~W.}} \AND
  \bauthor{\bsnm{Manski},~\bfnm{Charles~F}\binits{C.~F.}}
(\byear{2004}).
\btitle{Confidence intervals for partially identified parameters}.
\bjournal{Econometrica}
\bvolume{72}
\bpages{1845--1857}.
\end{barticle}
\endbibitem

\bibitem[\protect\citeauthoryear{Jackson}{1997}]{jackson1997compositional}
\begin{barticle}[author]
\bauthor{\bsnm{Jackson},~\bfnm{Donald~A}\binits{D.~A.}}
(\byear{1997}).
\btitle{Compositional data in community ecology: the paradigm or peril of
  proportions?}
\bjournal{Ecology}
\bvolume{78}
\bpages{929--940}.
\end{barticle}
\endbibitem

\bibitem[\protect\citeauthoryear{Kurtz et~al.}{2015}]{kurtz2015sparse}
\begin{barticle}[author]
\bauthor{\bsnm{Kurtz},~\bfnm{Zachary~D}\binits{Z.~D.}},
  \bauthor{\bsnm{M{\"u}ller},~\bfnm{Christian~L}\binits{C.~L.}},
  \bauthor{\bsnm{Miraldi},~\bfnm{Emily~R}\binits{E.~R.}},
  \bauthor{\bsnm{Littman},~\bfnm{Dan~R}\binits{D.~R.}},
  \bauthor{\bsnm{Blaser},~\bfnm{Martin~J}\binits{M.~J.}} \AND
  \bauthor{\bsnm{Bonneau},~\bfnm{Richard~A}\binits{R.~A.}}
(\byear{2015}).
\btitle{Sparse and compositionally robust inference of microbial ecological
  networks}.
\bjournal{PLoS Computational Biology}
\bvolume{11}
\bpages{e1004226}.
\end{barticle}
\endbibitem

\bibitem[\protect\citeauthoryear{Lin and Peddada}{2020}]{lin2020analysis}
\begin{barticle}[author]
\bauthor{\bsnm{Lin},~\bfnm{Huang}\binits{H.}} \AND
  \bauthor{\bsnm{Peddada},~\bfnm{Shyamal~Das}\binits{S.~D.}}
(\byear{2020}).
\btitle{Analysis of microbial compositions: a review of normalization and
  differential abundance analysis}.
\bjournal{NPJ biofilms and microbiomes}
\bvolume{6}
\bpages{1--13}.
\end{barticle}
\endbibitem

\bibitem[\protect\citeauthoryear{Love, Huber and
  Anders}{2014}]{love2014moderated}
\begin{barticle}[author]
\bauthor{\bsnm{Love},~\bfnm{Michael~I}\binits{M.~I.}},
  \bauthor{\bsnm{Huber},~\bfnm{Wolfgang}\binits{W.}} \AND
  \bauthor{\bsnm{Anders},~\bfnm{Simon}\binits{S.}}
(\byear{2014}).
\btitle{Moderated estimation of fold change and dispersion for {RNA-seq} data
  with {DESeq2}}.
\bjournal{Genome {B}iology}
\bvolume{15}
\bpages{1--21}.
\end{barticle}
\endbibitem

\bibitem[\protect\citeauthoryear{Lovell, Chua and
  McGrath}{2020}]{lovell2020counts}
\begin{barticle}[author]
\bauthor{\bsnm{Lovell},~\bfnm{David~R}\binits{D.~R.}},
  \bauthor{\bsnm{Chua},~\bfnm{Xin-Yi}\binits{X.-Y.}} \AND
  \bauthor{\bsnm{McGrath},~\bfnm{Annette}\binits{A.}}
(\byear{2020}).
\btitle{Counts: an outstanding challenge for log-ratio analysis of
  compositional data in the molecular biosciences}.
\bjournal{NAR Genomics and Bioinformatics}
\bvolume{2}
\bpages{lqaa040}.
\bdoi{10.1093/nargab/lqaa040}
\end{barticle}
\endbibitem

\bibitem[\protect\citeauthoryear{Lovell et~al.}{2011}]{lovell2011proportions}
\begin{barticle}[author]
\bauthor{\bsnm{Lovell},~\bfnm{David}\binits{D.}},
  \bauthor{\bsnm{M{\"u}ller},~\bfnm{Warren}\binits{W.}},
  \bauthor{\bsnm{Taylor},~\bfnm{Jen}\binits{J.}},
  \bauthor{\bsnm{Zwart},~\bfnm{Alec}\binits{A.}},
  \bauthor{\bsnm{Helliwell},~\bfnm{Chris}\binits{C.}},
  \bauthor{\bsnm{Pawlowsky-Glahn},~\bfnm{V}\binits{V.}} \AND
  \bauthor{\bsnm{Buccianti},~\bfnm{A}\binits{A.}}
(\byear{2011}).
\btitle{Proportions, percentages, ppm: do the molecular biosciences treat
  compositional data right}.
\bjournal{Compositional data analysis: Theory and applications}
\bpages{193--207}.
\end{barticle}
\endbibitem

\bibitem[\protect\citeauthoryear{Lovell
  et~al.}{2015}]{lovell2015proportionality}
\begin{barticle}[author]
\bauthor{\bsnm{Lovell},~\bfnm{David}\binits{D.}},
  \bauthor{\bsnm{Pawlowsky-Glahn},~\bfnm{Vera}\binits{V.}},
  \bauthor{\bsnm{Egozcue},~\bfnm{Juan~Jos{\'e}}\binits{J.~J.}},
  \bauthor{\bsnm{Marguerat},~\bfnm{Samuel}\binits{S.}} \AND
  \bauthor{\bsnm{B{\"a}hler},~\bfnm{J{\"u}rg}\binits{J.}}
(\byear{2015}).
\btitle{Proportionality: a valid alternative to correlation for relative data}.
\bjournal{PLoS Computational Biology}
\bvolume{11}
\bpages{e1004075}.
\end{barticle}
\endbibitem

\bibitem[\protect\citeauthoryear{McGovern, Nixon and
  Silverman}{2023}]{mcgovern2023addressing}
\begin{barticle}[author]
\bauthor{\bsnm{McGovern},~\bfnm{Kyle~C}\binits{K.~C.}},
  \bauthor{\bsnm{Nixon},~\bfnm{Michelle~Pistner}\binits{M.~P.}} \AND
  \bauthor{\bsnm{Silverman},~\bfnm{Justin~D}\binits{J.~D.}}
(\byear{2023}).
\btitle{{Addressing Erroneous Scale Assumptions in Microbe and Gene Set
  Enrichment Analysis}}.
\bjournal{PLoS Comput Biol}
\bvolume{19}.
\end{barticle}
\endbibitem

\bibitem[\protect\citeauthoryear{Murphy et~al.}{2016}]{murphy2016capture}
\begin{barticle}[author]
\bauthor{\bsnm{Murphy},~\bfnm{Sean~M.}\binits{S.~M.}},
  \bauthor{\bsnm{Cox},~\bfnm{John~J.}\binits{J.~J.}},
  \bauthor{\bsnm{Augustine},~\bfnm{Ben~C.}\binits{B.~C.}},
  \bauthor{\bsnm{Hast},~\bfnm{John~T.}\binits{J.~T.}},
  \bauthor{\bsnm{Guthrie},~\bfnm{Joseph~M.}\binits{J.~M.}},
  \bauthor{\bsnm{Wright},~\bfnm{John}\binits{J.}},
  \bauthor{\bsnm{McDermott},~\bfnm{Joseph}\binits{J.}},
  \bauthor{\bsnm{Maehr},~\bfnm{Sutton~C.}\binits{S.~C.}} \AND
  \bauthor{\bsnm{Plaxico},~\bfnm{Jayson~H.}\binits{J.~H.}}
(\byear{2016}).
\btitle{Characterizing recolonization by a reintroduced bear population using
  genetic spatial capture-recapture}.
\bjournal{The Journal of Wildlife Management}
\bvolume{80}
\bpages{1390–1407}.
\bdoi{10.1002/jwmg.21144}
\end{barticle}
\endbibitem

\bibitem[\protect\citeauthoryear{Nixon, Gloor and
  Silverman}{2024}]{nixon2024aldex2}
\begin{barticle}[author]
\bauthor{\bsnm{Nixon},~\bfnm{Michelle~Pistner}\binits{M.~P.}},
  \bauthor{\bsnm{Gloor},~\bfnm{Gregory~B.}\binits{G.~B.}} \AND
  \bauthor{\bsnm{Silverman},~\bfnm{Justin~D.}\binits{J.~D.}}
(\byear{2024}).
\btitle{Beyond {Normalization}: {Incorporating} {Scale} {Uncertainty} in
  {Microbiome} and {Gene} {Expression} {Analysis}}.
\bjournal{bioRxiv}.
\bdoi{10.1101/2024.04.01.587602}
\end{barticle}
\endbibitem

\bibitem[\protect\citeauthoryear{Pawlowsky-Glahn, Egozcue and
  Tolosana-Delgado}{2015}]{pawlowsky2015book}
\begin{bbook}[author]
\bauthor{\bsnm{Pawlowsky-Glahn},~\bfnm{Vera}\binits{V.}},
  \bauthor{\bsnm{Egozcue},~\bfnm{Juan~Jos{\'e}}\binits{J.~J.}} \AND
  \bauthor{\bsnm{Tolosana-Delgado},~\bfnm{Raimon}\binits{R.}}
(\byear{2015}).
\btitle{Modeling and analysis of compositional data}.
\bpublisher{John Wiley \& Sons}.
\end{bbook}
\endbibitem

\bibitem[\protect\citeauthoryear{Props et~al.}{2017}]{props2017absolute}
\begin{barticle}[author]
\bauthor{\bsnm{Props},~\bfnm{Ruben}\binits{R.}},
  \bauthor{\bsnm{Kerckhof},~\bfnm{Frederiek-Maarten}\binits{F.-M.}},
  \bauthor{\bsnm{Rubbens},~\bfnm{Peter}\binits{P.}},
  \bauthor{\bsnm{De~Vrieze},~\bfnm{Jo}\binits{J.}},
  \bauthor{\bsnm{Hernandez~Sanabria},~\bfnm{Emma}\binits{E.}},
  \bauthor{\bsnm{Waegeman},~\bfnm{Willem}\binits{W.}},
  \bauthor{\bsnm{Monsieurs},~\bfnm{Pieter}\binits{P.}},
  \bauthor{\bsnm{Hammes},~\bfnm{Frederik}\binits{F.}} \AND
  \bauthor{\bsnm{Boon},~\bfnm{Nico}\binits{N.}}
(\byear{2017}).
\btitle{Absolute quantification of microbial taxon abundances}.
\bjournal{The ISME journal}
\bvolume{11}
\bpages{584--587}.
\end{barticle}
\endbibitem

\bibitem[\protect\citeauthoryear{Roche and
  Mukherjee}{2022}]{roche2022composition}
\begin{barticle}[author]
\bauthor{\bsnm{Roche},~\bfnm{Kimberly~E.}\binits{K.~E.}} \AND
  \bauthor{\bsnm{Mukherjee},~\bfnm{Sayan}\binits{S.}}
(\byear{2022}).
\btitle{The accuracy of absolute differential abundance analysis from relative
  count data}.
\bjournal{PLOS Computational Biology}
\bvolume{18}
\bpages{e1010284}.
\bdoi{10.1371/journal.pcbi.1010284}
\end{barticle}
\endbibitem

\bibitem[\protect\citeauthoryear{Rossi et~al.}{1987}]{rossi1987urban}
\begin{barticle}[author]
\bauthor{\bsnm{Rossi},~\bfnm{Peter~H}\binits{P.~H.}},
  \bauthor{\bsnm{Wright},~\bfnm{James~D}\binits{J.~D.}},
  \bauthor{\bsnm{Fisher},~\bfnm{Gene~A}\binits{G.~A.}} \AND
  \bauthor{\bsnm{Willis},~\bfnm{Georgianna}\binits{G.}}
(\byear{1987}).
\btitle{The urban homeless: estimating composition and size}.
\bjournal{Science}
\bvolume{235}
\bpages{1336--1341}.
\end{barticle}
\endbibitem

\bibitem[\protect\citeauthoryear{Santos and Papa}{2022}]{santos2022avoiding}
\begin{barticle}[author]
\bauthor{\bsnm{Santos},~\bfnm{Claudio Filipi Gon{\c{c}}alves~Dos}\binits{C.~F.
  G.~D.}} \AND \bauthor{\bsnm{Papa},~\bfnm{Jo{\~a}o~Paulo}\binits{J.~P.}}
(\byear{2022}).
\btitle{{Avoiding overfitting: A survey on regularization methods for
  convolutional neural networks}}.
\bjournal{ACM Computing Surveys (CSUR)}
\bvolume{54}
\bpages{1--25}.
\end{barticle}
\endbibitem

\bibitem[\protect\citeauthoryear{Schwartz}{1965}]{schwartz1965bayes}
\begin{barticle}[author]
\bauthor{\bsnm{Schwartz},~\bfnm{Lorraine}\binits{L.}}
(\byear{1965}).
\btitle{On Bayes procedures}.
\bjournal{Zeitschrift f{\"u}r Wahrscheinlichkeitstheorie und verwandte Gebiete}
\bvolume{4}
\bpages{10--26}.
\end{barticle}
\endbibitem

\bibitem[\protect\citeauthoryear{Silverman, Hupert and
  Washburne}{2020}]{silverman2020ili}
\begin{barticle}[author]
\bauthor{\bsnm{Silverman},~\bfnm{Justin~D.}\binits{J.~D.}},
  \bauthor{\bsnm{Hupert},~\bfnm{Nathaniel}\binits{N.}} \AND
  \bauthor{\bsnm{Washburne},~\bfnm{Alex~D.}\binits{A.~D.}}
(\byear{2020}).
\btitle{Using influenza surveillance networks to estimate state-specific
  prevalence of {SARS}-{CoV}-2 in the {United} {States}}.
\bjournal{Science Translational Medicine}
\bvolume{12}
\bpages{eabc1126}.
\bdoi{10.1126/scitranslmed.abc1126}
\end{barticle}
\endbibitem

\bibitem[\protect\citeauthoryear{Silverman et~al.}{2018}]{silverman2018dlm}
\begin{barticle}[author]
\bauthor{\bsnm{Silverman},~\bfnm{Justin~D.}\binits{J.~D.}},
  \bauthor{\bsnm{Durand},~\bfnm{Heather~K.}\binits{H.~K.}},
  \bauthor{\bsnm{Bloom},~\bfnm{Rachael~J.}\binits{R.~J.}},
  \bauthor{\bsnm{Mukherjee},~\bfnm{Sayan}\binits{S.}} \AND
  \bauthor{\bsnm{David},~\bfnm{Lawrence~A.}\binits{L.~A.}}
(\byear{2018}).
\btitle{Dynamic linear models guide design and analysis of microbiota studies
  within artificial human guts}.
\bjournal{Microbiome}
\bvolume{6}.
\bdoi{10.1186/s40168-018-0584-3}
\end{barticle}
\endbibitem

\bibitem[\protect\citeauthoryear{Silverman
  et~al.}{2021}]{silverman2021measuring}
\begin{barticle}[author]
\bauthor{\bsnm{Silverman},~\bfnm{Justin~D}\binits{J.~D.}},
  \bauthor{\bsnm{Bloom},~\bfnm{Rachael~J}\binits{R.~J.}},
  \bauthor{\bsnm{Jiang},~\bfnm{Sharon}\binits{S.}},
  \bauthor{\bsnm{Durand},~\bfnm{Heather~K}\binits{H.~K.}},
  \bauthor{\bsnm{Dallow},~\bfnm{Eric}\binits{E.}},
  \bauthor{\bsnm{Mukherjee},~\bfnm{Sayan}\binits{S.}} \AND
  \bauthor{\bsnm{David},~\bfnm{Lawrence~A}\binits{L.~A.}}
(\byear{2021}).
\btitle{Measuring and mitigating {PCR} bias in microbiota datasets}.
\bjournal{PLoS Computational Biology}
\bvolume{17}
\bpages{e1009113}.
\end{barticle}
\endbibitem

\bibitem[\protect\citeauthoryear{Silverman
  et~al.}{2022}]{silverman2022bayesian}
\begin{barticle}[author]
\bauthor{\bsnm{Silverman},~\bfnm{Justin~D}\binits{J.~D.}},
  \bauthor{\bsnm{Roche},~\bfnm{Kimberly}\binits{K.}},
  \bauthor{\bsnm{Holmes},~\bfnm{Zachary~C}\binits{Z.~C.}},
  \bauthor{\bsnm{David},~\bfnm{Lawrence~A}\binits{L.~A.}} \AND
  \bauthor{\bsnm{Mukherjee},~\bfnm{Sayan}\binits{S.}}
(\byear{2022}).
\btitle{{Bayesian Multinomial Logistic Normal Models through Marginally Latent
  Matrix-T Processes}}.
\bjournal{Journal of Machine Learning Research}
\bvolume{23}
\bpages{1--42}.
\end{barticle}
\endbibitem

\bibitem[\protect\citeauthoryear{Srinivasan, Xue and
  Zhan}{2020}]{srinivasan2020knockoff}
\begin{barticle}[author]
\bauthor{\bsnm{Srinivasan},~\bfnm{Arun}\binits{A.}},
  \bauthor{\bsnm{Xue},~\bfnm{Lingzhou}\binits{L.}} \AND
  \bauthor{\bsnm{Zhan},~\bfnm{Xiang}\binits{X.}}
(\byear{2020}).
\btitle{Compositional knockoff filter for high-dimensional regression analysis
  of microbiome data}.
\bjournal{Biometrics}.
\bdoi{10.1111/biom.13336}
\end{barticle}
\endbibitem

\bibitem[\protect\citeauthoryear{Stoye}{2009}]{stoye2009more}
\begin{barticle}[author]
\bauthor{\bsnm{Stoye},~\bfnm{J{\"o}rg}\binits{J.}}
(\byear{2009}).
\btitle{More on confidence intervals for partially identified parameters}.
\bjournal{Econometrica}
\bvolume{77}
\bpages{1299--1315}.
\end{barticle}
\endbibitem

\bibitem[\protect\citeauthoryear{Top{\^\i}rceanu and
  Precup}{2020}]{topirceanu2020polling}
\begin{barticle}[author]
\bauthor{\bsnm{Top{\^\i}rceanu},~\bfnm{Alexandru}\binits{A.}} \AND
  \bauthor{\bsnm{Precup},~\bfnm{Radu-Emil}\binits{R.-E.}}
(\byear{2020}).
\btitle{A framework for improving electoral forecasting based on time-aware
  polling}.
\bjournal{Social Network Analysis and Mining}
\bvolume{10}
\bpages{39}.
\end{barticle}
\endbibitem

\bibitem[\protect\citeauthoryear{Vandeputte
  et~al.}{2017}]{vandeputte2017quantitative}
\begin{barticle}[author]
\bauthor{\bsnm{Vandeputte},~\bfnm{Doris}\binits{D.}},
  \bauthor{\bsnm{Kathagen},~\bfnm{Gunter}\binits{G.}},
  \bauthor{\bsnm{D’hoe},~\bfnm{Kevin}\binits{K.}},
  \bauthor{\bsnm{Vieira-Silva},~\bfnm{Sara}\binits{S.}},
  \bauthor{\bsnm{Valles-Colomer},~\bfnm{Mireia}\binits{M.}},
  \bauthor{\bsnm{Sabino},~\bfnm{Jo{\~a}o}\binits{J.}},
  \bauthor{\bsnm{Wang},~\bfnm{Jun}\binits{J.}},
  \bauthor{\bsnm{Tito},~\bfnm{Raul~Y}\binits{R.~Y.}},
  \bauthor{\bsnm{De~Commer},~\bfnm{Lindsey}\binits{L.}},
  \bauthor{\bsnm{Darzi},~\bfnm{Youssef}\binits{Y.}} \betal{et~al.}
(\byear{2017}).
\btitle{Quantitative microbiome profiling links gut community variation to
  microbial load}.
\bjournal{Nature}
\bvolume{551}
\bpages{507--511}.
\end{barticle}
\endbibitem

\bibitem[\protect\citeauthoryear{Yu}{2020}]{yu2020veridical}
\begin{binproceedings}[author]
\bauthor{\bsnm{Yu},~\bfnm{Bin}\binits{B.}}
(\byear{2020}).
\btitle{Veridical data science}.
In \bbooktitle{Proceedings of the 13th International Conference on Web Search
  and Data Mining}
\bpages{4--5}.
\end{binproceedings}
\endbibitem

\end{thebibliography}


\begin{thebibliography}{}

\bibitem[Fernandes et~al., 2014]{fernandes2014unifying}
Fernandes, A.~D., Reid, J.~N., Macklaim, J.~M., McMurrough, T.~A., Edgell,
  D.~R., and Gloor, G.~B. (2014).
\newblock Unifying the analysis of high-throughput sequencing datasets:
  characterizing {RNA-seq}, {16S} {rRNA} gene sequencing and selective growth
  experiments by compositional data analysis.
\newblock {\em Microbiome}, 2(1):1--13.

\bibitem[Gabrielsen, 1978]{gabrielsen1978consistency}
Gabrielsen, A. (1978).
\newblock Consistency and identifiability.
\newblock {\em Journal of Econometrics}, 8(2):261--263.

\bibitem[Gupta and Nagar, 2000]{gupta2000matrix}
Gupta, A. and Nagar, D. (2000).
\newblock {\em Matrix Variate Distributions}.
\newblock Chapman \& Hall/CRC.

\bibitem[Krebs, 1999]{krebs1999book}
Krebs, C.~J. (1999).
\newblock {\em Ecological Methodology}.
\newblock Benjamin/Cummings, Menlo Park, Calif, 2nd ed edition.

\bibitem[Letourneau et~al., 2022]{letourneau2021ecological}
Letourneau, J., Holmes, Z.~C., Dallow, E.~P., Durand, H.~K., Jiang, S.,
  Carrion, V.~M., Gupta, S.~K., Mincey, A.~C., Muehlbauer, M.~J., Bain, J.~R.,
  et~al. (2022).
\newblock Ecological memory of prior nutrient exposure in the human gut
  microbiome.
\newblock {\em The ISME Journal}, 16(11):2479--2490.

\bibitem[Love et~al., 2014]{love2014moderated}
Love, M.~I., Huber, W., and Anders, S. (2014).
\newblock Moderated estimation of fold change and dispersion for {RNA-seq} data
  with {DESeq2}.
\newblock {\em Genome {B}iology}, 15(12):1--21.

\bibitem[Oosterhoff, 1969]{oosterhoff1969combination}
Oosterhoff, J. (1969).
\newblock {\em Combination of one-sided statistical tests}.
\newblock Mathematical Centre, Amsterdam.

\bibitem[Pawlowsky-Glahn et~al., 2015]{pawlowsky2015book}
Pawlowsky-Glahn, V., Egozcue, J.~J., and Tolosana-Delgado, R. (2015).
\newblock {\em Modeling and analysis of compositional data}.
\newblock John Wiley \& Sons.

\bibitem[Silverman et~al., 2018]{silverman2018dynamic}
Silverman, J.~D., Durand, H.~K., Bloom, R.~J., Mukherjee, S., and David, L.~A.
  (2018).
\newblock Dynamic linear models guide design and analysis of microbiota studies
  within artificial human guts.
\newblock {\em Microbiome}, 6(1):1--20.

\bibitem[Silverman et~al., 2022]{silverman2022bayesian}
Silverman, J.~D., Roche, K., Holmes, Z.~C., David, L.~A., and Mukherjee, S.
  (2022).
\newblock {Bayesian Multinomial Logistic Normal Models through Marginally
  Latent Matrix-T Processes}.
\newblock {\em Journal of Machine Learning Research}, 23(7):1--42.

\bibitem[Srinivasan et~al., 2020]{srinivasan2020knockoff}
Srinivasan, A., Xue, L., and Zhan, X. (2020).
\newblock Compositional knockoff filter for high-dimensional regression
  analysis of microbiome data.
\newblock {\em Biometrics}.

\bibitem[Udevitz and Pollock, 1992]{udevitz1992change}
Udevitz, M.~S. and Pollock, K.~H. (1992).
\newblock Change-in-ratio methods for estimating population size.
\newblock In {\em Wildlife 2001: Populations}, pages 90--101. Springer.

\bibitem[Vandeputte et~al., 2017]{vandeputte2017quantitative}
Vandeputte, D., Kathagen, G., D’hoe, K., Vieira-Silva, S., Valles-Colomer,
  M., Sabino, J., Wang, J., Tito, R.~Y., De~Commer, L., Darzi, Y., et~al.
  (2017).
\newblock Quantitative microbiome profiling links gut community variation to
  microbial load.
\newblock {\em Nature}, 551(7681):507--511.

\end{thebibliography}

\end{document}


\def\spacingset#1{\renewcommand{\baselinestretch}%
  {#1}\small\normalsize} \spacingset{1}


\if1\blind { \title{\bf Supplement to ``Scale Reliant Inference''} \author{Michelle Pistner
    Nixon,
    Kyle C. McGovern,
    Jeffrey Letourneau,\\
    Lawrence A. David, 
    Nicole A. Lazar,
    Sayan Mukherjee,
    and\\
    Justin D. Silverman}  \maketitle
} \fi

\if0\blind {
    {\title{\bf Supplement to ``Scale Reliant Inference''}
    \maketitle}
   } \fi


\section{Toy Example}

In Figure~\ref{fig:toy-example}, we provide a toy example showing the true abundances \(\W\) and the sequence count table \(\Y\) for a study of \(N=3\) microbial communities each consisting of \(D=4\) taxa. Elements \(Y_{dn}\) denote the number of DNA molecules sequenced (measured) in a fecal sample from subject \(n\) that are from taxon \(d\). Elements \(W_{dn}\) represent the true amount of taxon \(d\) in the gut of subject \(n\).

This table introduces the notation \(\YC\) and \(\YT\) which are the analogs of
\(\WC\) and \(\WT\): \(\YT_{n}=\sum_{d} \Y_{dn}\) and \(\YC_{dn}=\Y_{dn}/\YT_{n}\). We highlight that the composition of the system \(\WC\) and the data \(\YC\) are approximately equal (up to sampling noise) whereas the scale of the system \(\WT\) and \(\YT\) may be unrelated. To highlight this point, we have designed this example such that \(\YT_{2} >\WT_{2}\) (more measured than actually present). In microbiome data, this could occur due to the polymerase chain reaction (PCR) part of the measurement process. 

As stated in Section~2 of the main text, knowledge of the system scale alone (\(\WC\)) is insufficient to estimate log-fold-changes
\[\theta_{d} = \mean_{n:x_{n}=1}\log \W_{dn}-\mean_{n:x_{n}=0}\log \W_{dn}.\]
From main text,  \(\W_{dn}=\WC_{dn}\WT_{n}\) and therefore 
\begin{align}
\theta_{d}&=\underbrace{\mean_{n:x_{n}=1}(\log \WC_{dn}) -\mean_{n:x_{n}=0}(\log \WC_{dn})}_{\thetaC_{d}}+ \underbrace{\mean_{n:x_{n}=1}(\log \WT_{n}) -\mean_{n:x_{n}=0}(\log \WT_{n})}_{\thetaT} \nonumber  \\ 
 &= \thetaC_{d} + \thetaT. \label{eq:lfc-decomposition}
\end{align}

To illustrate the problem, suppose a researchers knows (without error) the value of \(\thetaC_{d}>0\). Without knowledge of \(\thetaT\), that researcher cannot even determine if \(\theta_{d}\) is equal to, greater than, or less than zero. In our toy example, \(\theta=(0, 0.40, 0.55, 0.69)\), \(\thetaC=(-0.45, -0.05, 0.10, 0.25)\) and \(\thetaT=0.45\). Note that, the sign of \(\thetaC_{1}\) and \(\thetaC_{2}\) is flipped compared to \(\theta_{1}\) and \(\theta_{2}\). This highlights that using $\thetaC$ to form conclusions about $\theta$ can lead to incorrect inferences.

\begin{figure}[H]
  \centering
  \begin{tabular}{|c|ccc |}
    \hline
    \(\W\) {\color{gray}(\(\WC\))}& \(n=1\) & \(n=2\) & \(n=3\) \\
    \hline
    \(d=1\)& 10 {\color{gray} (0.10)} & 10 {\color{gray} (0.08)} & 10 {\color{gray} (0.14)} \\
    \(d=2\)& 40 {\color{gray} (0.40)} & 50 {\color{gray} (0.42)} & 30 {\color{gray} (0.43)}  \\
    \(d=3\)& 30 {\color{gray} (0.30)} & 40 {\color{gray} (0.33)} & 20 {\color{gray} (0.29)}  \\
    \(d=4\)& 20 {\color{gray} (0.20)} & 20 {\color{gray} (0.17)} & 10 {\color{gray} (0.14)} \\
    \hline\hline
     {\color{gray}\(\WT\)}& {\color{gray} 100} & {\color{gray} 120} & {\color{gray} 70} \\
    \hline
  \end{tabular}
  \;
  \begin{tabular}{|c|ccc|}
    \hline
    \(\Y\) {\color{gray}(\(\YC\))}& \(n=1\) & \(n=2\) & \(n=3\) \\
    \hline
    \(d=1\)& 3 {\color{gray} (0.15)} & 14 {\color{gray} (0.07)} & 15 {\color{gray} (0.15)} \\
    \(d=2\)& 7 {\color{gray} (0.35)} & 95 {\color{gray} (0.48)} & 35 {\color{gray} (0.35)}  \\
    \(d=3\)& 7 {\color{gray} (0.35)} & 65 {\color{gray} (0.32)} & 30 {\color{gray} (0.30)}  \\
    \(d=4\)& 3 {\color{gray} (0.15)} & 26 {\color{gray} (0.13)} & 20 {\color{gray} (0.20)} \\
    \hline\hline
     {\color{gray}\(\YT\)}& {\color{gray} 20} & {\color{gray} 200} & {\color{gray} 100} \\
    \hline
  \end{tabular}
  \par\bigskip
  \begin{tabular}{|c|ccc|}
    \hline
    & \(n=1\) & \(n=2\) & \(n=3\) \\
    \hline
    \(x\) & 1 & 1 & 0 \\
    \hline
  \end{tabular}
  \caption{A toy microbiome study. \(\mathbf{W}\) denotes the unmeasured system, \(\Y\) the observed data, and \(x\) a single covariate distinguishing two biological conditions (e.g., health versus disease). We have created this toy example such that the only taxon that is differentially abundant between these two conditions is taxon \(d=1\). Still, the arbitrary scale of the data distort this picture and can leave naive analyses to spurious conclusions.} 
  \label{fig:toy-example}

\end{figure}

\section{Proofs}

\subsection{Log-Scale Covariance is a Partially Scale Reliant Estimand}
\label{sec:covariance}

\begin{proposition}
  \(\Sigma_{d_{1}, d_{2}} = cov(\log \W_{d_{1} \cdot}, \log \W_{d_{2} \cdot})\) is partially scale reliant so long as \(cov(\log \WC_{d_{1} \cdot}, \log \WC_{d_{2} \cdot})\) is finite and \(var(\log \WT)\) is finite. 
\end{proposition}
\begin{proof}
  For notational convenience, let \(\logW\) be a shorthand for \(\log \W\). 
  We reparameterize \(\Sigma_{d_1d_2}\) in terms of \(\logWCBF\) and \(\logWT\). 
  \begin{equation*}
    \begin{split}
      \Sigma_{d_1d_2} &= \text{cov}{(\logWC_{d_1 \cdot} + \logWT, \logWC_{d_2 \cdot} + \logWT)}\\
      &= \text{cov}(\logWC_{d_1 \cdot}, \logWC_{d_2 \cdot}) + \text{cov}(\logWC_{d_1 \cdot}, \logWT)
      + \text{cov}(\logWT, \logWC_{d_2 \cdot}) + \text{cov}(\logWT, \logWT)\\
    \end{split}
  \end{equation*}
  We denote the standard deviation of a vector \(x\) as \(\sigma_{x}=\sqrt{\text{var}(x)}\)
  and the covariation of two vectors \(x\) and \(y\) as \(\rho_{xy} = corr(x, y)\). We can
  then write:
  \begin{align*}
    \Sigma_{d_1d_2} &= \sigma_{\logWC_{d_1 \cdot}}\sigma_{\logWC_{d_2 \cdot}} \rho(\logWC_{d_1 \cdot}, \logWC_{d_2 \cdot})+ \sigma_{\logWC_{d_1 \cdot}}\sigma_{\logWT}\rho(\logWC_{d_1 \cdot}, \logWT) \\
    & \, \, \, \, \, \,  + \sigma_{\logWC_{d_2 \cdot}}\sigma_{\logWT}\rho(\logWC_{d_2 \cdot}, \logWT) + \sigma_{\logWT}^{2} \\
 &= \alpha + \sigma_{\logWT}\beta + \sigma_{\logWT}\gamma + \sigma^{2}_{\logWT}. 
  \end{align*}
  For a fixed value of \(\alpha\), the minimum value that \(\Sigma_{d_{1}d_{2}}\) can attain is \(\alpha-(\beta+\gamma)^{2}/4\) which occurs when \(\sigma^{2}_{\logWT}=(\beta+\gamma)/2\). 
  Note that as long as \(\alpha=\text{cov}(\logWC_{d_1 \cdot}, \logWC_{d_2 \cdot})\) is finite and \(\sigma_{\overline{W}^{tot}}\) is finite, then \(\beta\) and \(\gamma\) must also be finite. It follows that the estimand is therefore partially scale reliant so long as \(\text{cov}(\logWC_{d_1 \cdot}, \logWC_{d_2 \cdot})\) is finite and \(\text{var}(\logWT)\) is finite. 
\end{proof}

\subsection{Scale Reliance in Relation to Estimands and Compositional Surveys}
\label{sec:scale-reliant-inference-results}

In this section, we establish the relationship between  Scale Invariant Inference (SII)/Scale Reliant Inference (SRI) and  various compositional survey and target estimand pairings. To simplify our presentation, in this section, when we say a target estimand is scale invariant/reliant, we mean that it is scale invariant/reliant at the particular \(\WC\) associated with the compositional survey \(\Y\). 

\begin{table}[H]
  \centering
  \caption{Inference of $\theta$ depending on the type of compositional survey (partial or full) and target estimand (scale invariant or scale reliant).}
  \begin{tabular}{c c c c}
    \hline
     Compositional Survey (\(\Y\)) & Target Estimand (\(\theta\))& Inference of \(\theta\) from \\
 & &  observations of \(\Y\)\\
    \hline 
    Full    & Scale Invariant         &  Scale Invariant     \\
    Full    & Partially Scale Reliant &  Partially Scale Reliant \\
    Full    & Fully Scale Reliant     &  Fully Scale Reliant \\
    Partial & Scale Invariant         &  Scale Invariant     \\
    Partial & Partially Scale Reliant &  --     \\
    Partial & Fully Scale Reliant     &  --     \\
    \hline 
  \end{tabular}\label{tab:scale-reliant-inference-results}
\end{table}

To prove the results of this table, we will need the following lemma which states that \(\Id{\theta} P\) is equal to the image of the identification region \(\Id{(\WC, \WT)} P\) under the target estimand. 

\begin{lemma}
  If \(\Y\) is a compositional survey and \(\theta\) a target estimand, then
\[\Id {\theta} P=\{\theta\in \Theta: \exists \; (\WC, \WT)\in \Id {\WC, \WT} P \text{ s.t. } \theta=\theta(\WC, \WT)\}.\] 
\end{lemma}
\begin{proof}
  This follows from the fact that \(\theta\) is a function of \(\WC\) and \(\WT\). 
\end{proof}

The results in the table follow immediately from the lemma. First, we use the lemma to establish the first three rows of the table. Consider that for a full compositional survey \(Y \sim P\), \(\Id {\WC, \WT} P\) is equal to \(\WC\times \WTS\). This is precisely the same set that we used to define whether a target estimand is scale invariant/reliant in the previous section. Based on the above lemma, it follows that \(\Id {\theta} P=\theta(\WC \times \WTS)\). It follows that if \(\theta\) is scale invariant, then \(\Id {\theta} P\) is a singleton set (SII). The same logic establishes that if \(\theta\) is partially/fully scale reliant then the problem is Partial/Full SRI. 

The fourth row of the above table follows directly from the fact that \(\WC\) is identified in a partial compositional survey (\(\Id {\WC} P\) is a singleton set) in combination with the above lemma which establishes that, if \(\theta\) is scale invariant, than a single element \(\WC\in \Id{\WC} P\) implies a single element in \(\Id{\theta} P\) (SII). 

 There are no simple results for the fifth and sixth rows. Whether a problem is SII, Full SRI, or Partial SRI depends on the specific features of the estimands and surveys in question beyond those captured in the categorization we have discussed so far. We leave further exploration of these cases to future research.  

\subsection{Proofs for Limits on Frequentist Inference}
\label{sec:proofs-limits}

Let \(\{\hat{\theta}_{n}\}\) denote a sequence of estimators for the target estimand \(\theta\). 
\begin{definition}[Consistent Estimator]
  \(\{\hat{\theta}_{n}\}\) is a consistent sequence of estimators for \(\theta\) if and only if, for all \(\epsilon>0\) 
  \[\lim_{n \rightarrow \infty} P(||\hat{\theta}_{n} - \theta|| < \epsilon) = 1.\]
\end{definition}

The following proof is a straightforward adaptation of standard results stating that identifiability is a necessary condition for consistency (e.g., \cite{gabrielsen1978consistency}). 
\begin{theorem}
  (Consistency) If \(\Id {\theta} P\) is not a singleton set (Full or Partial SRI), then there does not exist a sequence of estimators for the target estimand \(\theta\) that is consistent. 
\end{theorem}
\begin{proof}
  In the setting of Full or Partial SRI, \(\Id {\theta} P\) must contain at least two distinct elements \(\theta_{1}, \theta_{2}\). By contradiction, suppose that there does exist a sequence of estimators \(\{\hat{\theta}_{n}\}\) for a Partial or Full SRI problem that is consistent. Then, the sequence must asymptotically converge to both \(\theta_{1}\) and \(\theta_{2}\) simultaneously. Yet \(\theta_{1}\neq \theta_{2}\). So convergence to both \(\theta_{1}\) and \(\theta_{2}\) (to arbitrary \(\epsilon\) precision) is impossible. 
\end{proof}

\begin{definition}[Unbiased Estimator]
  The bias of an estimator \(\hat{\theta}=\hat{\theta}(\Y)\) where \(\Y \sim P_{\theta, \lambda}\in \mathbf{P}\) is
  \[\text{bias}_{P}(\hat{\theta})=\mathbb{E}_{P}[\hat{\theta}]-\theta.\] 
  An estimator \(\hat{\theta}\) is unbiased if \(\text{bias}_{P}(\hat{\theta}) =0\). 
\end{definition}

\begin{theorem}
  (Unbiasedness) If \(\Id {\theta} P\) is not a singleton set (Full or Partial SRI), then there does not exist an estimator for the target estimand \(\theta\) that is unbiased. 
\end{theorem}
\begin{proof}
  In the setting of Full or Partial SRI, \(\Id {\theta} P\) must contain at least two distinct elements \(\theta_{1}, \theta_{2}\). By contradiction, suppose that there does exist an unbiased estimator \(\hat{\theta}\) in this setting. Then, by definition, \(\mathbb{E}_{P}[\hat{\theta}]=\theta_{1}\) and \(\mathbb{E}_{P}[\hat{\theta}]=\theta_{2}\). Since \(\theta_{1}\neq \theta_{2}\), this is a contradiction. 
\end{proof}

\begin{definition}[Confidence Set]
  A set estimator \(S(\Y)\subset \Theta\) is calibrated at level \(1-\alpha\) if
  \[\inf_{\theta\in \Theta, \lambda \in \Lambda} P_{\theta, \lambda}(\theta \in S)\geq 1-\alpha.\] 
  The confidence set is non-trivial if, for each pair \((\theta, \lambda)\), there exists some \(\theta^{*}\in \Theta\) such that \(P_{\theta, \lambda}(\theta^{ * } \in S)< 1-\alpha\). 
\end{definition}
The triviality requirement essentially states that a confidence set cannot trivially cover the truth with probability \(1-\alpha\) by simply covering all elements with probability \(1-\alpha\). 

\begin{theorem}
  (Calibration) Suppose there exists \(\theta\in \Theta\) and \(\lambda\in \Lambda\) such that \(\Id {\theta} {P} =\Theta\) (Full SRI). 
  There does not exist a non-trivial confidence set \(S(\Y)\) calibrated at level \mbox{\(1-\alpha>0\)}. 
\end{theorem}
\begin{proof}
  By contradiction, assume that \(S(\Y)\) is a non-trivial confidence set calibrated at level \(1-\alpha>0\). Let \((\theta, \lambda)\) denote one of the multiple pairs at which \(\Id {\theta}{P_{\theta, \lambda}(\Y)}=\Theta\). As a non-trivial confidence set, then there must be some \(\theta^{*}\in \Theta\) such that \(P_{\theta, \lambda}(\theta^{*} \in S(\Y))< 1-\alpha\). However, because \(\Id {\theta} {P_{\theta, \lambda}(\Y)}=\Theta\), for \(S(\Y)\) to be calibrated at level \(1-\alpha\), \(S(\Y)\) must cover all elements in \(\Theta\) with probability at least \(1-\alpha\). So long as \(1-\alpha>0\), this is a contradiction. 
\end{proof}

An intriguing feature of this result is that: if in a family of distributions \(\mathbf{P}\) there exists even a single distribution at which the estimation problem is Full SRI, then there cannot exist any non-trivial confidence sets. This result is implied by the infimum over all \(\theta\in \Theta, \lambda\in \Lambda\) in the standard definition of confidence sets. Yet for some problems, this may not be useful as the data may strongly suggest that the true distribution \(P\) is not such a problematic case. Further studying this feature and developing set estimators that can take advantage of this intuition would require relaxing the standard definition of confidence sets. We leave such relaxations for future work.

\begin{definition}
  Let \(T(\Y)\) denote a test of the null hypothesis \(H_{0}:\theta\in \Theta_{0},\lambda\in \Lambda \). Let \(\beta(\theta, \lambda)\) denote the power of the test for a specific \(P_{\theta, \lambda}=P\in \mathbf{P}\): \(\beta(\theta, \lambda)=P_{\theta, \lambda}(T(\Y)=1)\). \(T(\Y)\) is a level-\(\alpha\) test if
  \[\sup_{\theta \in \Theta_{0}, \lambda\in \Lambda}\beta(\theta, \lambda)\leq \alpha.\]
  The test is non-trivial if there exist at least one \(\theta^{*} \notin \Theta_{0}, \lambda \in \Lambda\) such that \(\beta(\theta^{*}, \lambda)>\alpha\). 
\end{definition}
The non-trivial condition essentially states that the test has power greater than \(\alpha\) for at least one value of \(\theta\) within the alternative. We add the non-trivial condition to exclude trivial tests that, for example, never reject the null hypothesis. 

\begin{theorem}
  (Type-I Error Control) Suppose there exists some \(\theta\in \Theta\) and \(\lambda \in \Lambda\) such that \(\Id {\theta} {P}=\Theta\) (Full SRI). Then, for any \(\alpha>0\), there does not exist a non-trivial level-\(\alpha\) test of the null hypothesis \(\theta\in \Theta_{0}\subsetneq \Theta\).
  \label{thm:fdr}
\end{theorem}
\begin{proof}
  By contradiction, suppose such a tests, \(T(\Y)\), exists. Let \((\theta, \lambda)\) denote one of the multiple potential pairs at which \(\Id {\theta}{P_{\theta, \lambda}(\Y)}=\Theta\). As a non-trivial test, there must be some \(\theta^{*} \notin \Theta_{0}\) and \(\lambda\in \Lambda\) such that \(\beta(\theta^{*}, \lambda)>\alpha\). However, since \(\Id {\theta}{P_{\theta, \lambda}(\Y)}=\Theta\), it follows that to be a level-\(\alpha\) test \(\beta(\theta, \lambda)\leq \alpha\) for all \(\theta\in \Theta, \lambda \in \Lambda\). For any \(\alpha>0\), this is a contradiction. 
\end{proof}

\section{Review of Compositional Differential Abundance}
\label{sec:compositional-DA}

There are a number of articles that claim to develop hypothesis tests for differential abundance analysis for compositional microbiome data with type-I error control (e.g., \cite{srinivasan2020knockoff}). As discussed in the main text, these claims should be viewed skeptically: Theorem \ref{thm:fdr} in combination with Row 3 of Table \ref{tab:scale-reliant-inference-results} proves the only tests that control type-I error when estimating LFCs (a fully scale reliant estimand) from compositional data (full compositional surveys) are trivial: they never have power greater than \(\alpha\). Given that these articles claim to be able to detect true signal (e.g., non-trivial power), there is an apparent disagreement between our results and those claims. There are two possible explanations that can reconcile this disagreement; both suggest caution in interpreting the claims of those articles.

First, we provide an example of such an article. Recently, \cite{srinivasan2020knockoff} proposed a knockoff filter which they claim provides strict FDR control in microbiome DA analyses. The authors claim the data is compositional. We use our notation to describe their model. Let \(z\in \mathbb{R}^{N}\) denote a response vector and \(\tilde{\mathbf{Y}}\in \mathbb{R}^{D \times N}\) denote a center log-ratio normalized matrix of the count data $\Y$. They propose a regression model of the form:
\[z_{n}=\sum_{d=1}^{D}\tilde{\mathbf{Y}}_{dn}\beta_{d} + \epsilon_{n}\]
where they impose the identifying restriction \(\sum_{d}\beta_{d}=0\) for identifiability. The authors ultimately develop a knock-off filter to test, for each \(d\in \{1, \dots, D\}\), the null hypothesis \(H_{0}:\beta_{d}=0\). The authors state that a microbe \(d\) is differentially abundant if their test rejects this null hypothesis. 


The first explanation for the apparent disagreement between our results and their claims is that they are simply operating under a different estimand (a different definition for what it means for a microbe to be differentially abundant). Since the authors do not distinguish between their model and their scientific question (e.g., the thing they are trying to estimate), it is impossible for us to ascertain if this explanation is true or not. Still, we can explore what type of estimand is implied by their model and suggest that this estimand is unlikely to be of widespread interest to biomedical researchers. Consider that since \(\mathbf{Y}\) is essentially a log-transformed composition and \(\sum_{d}\beta_{d}=0\), the elements of \(\beta\) can be thought of as CLR coordinates. To take some minor liberties in our presentation for the sake of clarity (e.g., treating \(z\) as a binary variable, 
 and assuming no measurement noise in the observations \(\tilde{\mathbf{Y}}\) such that \(\tilde{\mathbf{Y}}\) is equal to the CLR transform of \(\WC\)), it is almost as if the authors are not defining log-fold-changes as we did in Equation 1 in the main text but instead as:
\[\beta_{d}=\mean_{n:x_{n}=1}(\tilde{\mathbf{W}}_{dn})-\mean_{n:x_{n}=0}(\tilde{\mathbf{W}}_{dn})\]
where \(\tilde{\W}_{dn}=\log \WC_{dn}- \mean_{d}(\log \WC_{dn})\) are CLR coordinates. Numerous authors in the CoDA literature have warned against trying to interpret CLR based quantities like this~\citep{pawlowsky2015book}. In short, \(\beta_{d}\) is not capturing the change in the abundance of taxon \(d\) in the way that biomedical researchers want but instead is representing how the abundance of taxon \(d\) changes relative to the mean change of all the other taxa. If one taxon increases in abundance, the mean changes as well, and all other taxa may end up looking like they have decreased (relative to the mean). To summarise, one explanation is that the authors are operating under a different definition of differential abundance; a definition that is difficult to interpret and has been warned against. 

The second explanation is that the authors are interested in the more biomedically relevant LFC estimand (e.g., Equation 2 in the main text) and that their claims of Type-I error control are erroneous. If the authors are interested in an estimand such as Equation 2 in the main text, then their null hypothesis is misspecified, and the theory surrounding knock-offs is insufficient to provide guarantees of FDR control. In fact, in this case, the only evidence that is relevant for claims of FDR control is the authors' simulations. Notably, their simulation scheme only simulated from their model -- the only simulated values of \(\beta\) that satisfied the identifying restriction \(\sum_{d}\beta_{d}=0\). As a result, this simulation scheme was insufficient to establish FDR control with respect to the more general LFC estimand in Equation 1 in the main text where there is no restriction that \(\sum_{d}\beta_{d}=0\). 

\section{Pitfalls in Specifying Bayesian Models for SRI}
\label{sec:bayesian-pitfalls}

\begin{figure}[ht]
  \centering
  \includegraphics[width=2in]{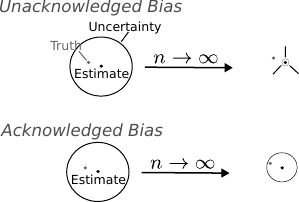}
  \caption{\footnotesize The relationship between unacknowledged bias and sample size can be
    understood as a symptom of model misspecification. Typical model-based estimators demonstrate
    decreasing uncertainty (e.g., decreasing variance) with increasing data, approaching zero in the
    asymptotic limit. Now consider a model that is misspecified such that a corresponding estimator
    displays a large-sample bias. When the sample size is small, the uncertainty can be large and
    encapsulate the bias: large uncertainty arising from even a misspecified model may buffer
    against the inferential effects of the bias. Yet with increasing sample size, the model becomes
    increasingly certain in a biased estimate; in turn, the impact of the bias becomes magnified. In
    the present context, model misspecification refers to models failing to incorporate uncertainty
    from the lack of scale information in the observed data.}
  \label{fig:unacknowledged-bias-intuition}
\end{figure}

In this section, we discuss some common pitfalls that can lead to unacknowledged bias. For an intuitive description of unacknowledged bias, see Figure \ref{fig:unacknowledged-bias-intuition}.
\subsection{Pitfall 1: Identified Likelihoods through Assumptions of Conditional Independence}

It is often tempting to model \(\Y\) as \(D \times N\) conditionally independent univariate measurements. For example,  in estimation of LFCs (\(\theta_{d}\)) from sequence count data, many models suppose that the elements \(\Y_{dn}\) are conditionally independent and described by either Poisson or Negative Binomial models, e.g.,  
\begin{align*}
\Y_{dn} &\sim \text{Poisson}(\W_{dn}) \\ 
\W_{dn} &= \exp(\beta_{d}+\theta_{d}x_{n}). 
\end{align*}
Asymptotically, the parameters \(\theta_{d}\) are identified in this model, leading to unacknowledged bias. 

\subsection{Pitfall 2: Identified Likelihoods through Deterministic Imputation}
Some models attempt to impute a scaling factor \(\mathbf{S}_{dn}\) that connects the scale of \(\Y\) to the scale of \(\W\), often a deterministic function \(\mathbf{S}_{dn}=f_{dn}(\Y)\). For example, the popular DESeq2 package for sequence count data analysis \citep{love2014moderated} uses a likelihood model of the form 
\begin{align*}
\Y_{dn} &\sim \text{NegBinomial}(\mathbf{S}_{dn}\W_{dn}, \alpha_{d}) \\ 
\W_{dn} &= \exp(\beta_{d}+\theta_{d}x_{n})
\end{align*}
where, by default \(\mathbf{S}_{dn}=S_{n}\) for all \(d\in \{1,\dots, D\}\) and \(S_{n}\) is defined by the following median-of-ratios calculation:
\begin{align*}
K_{d}^{R}&=G(\Y_{d\cdot}) \\ 
S_{n}&=\underset{d:K_{d}^{R}\neq 0}{\text{median }} \Y_{dn}/K_{d}^{R}. 
\end{align*}
Since there is not uncertainty considered in the relationship between \(\mathbf{S}_{dn}\) and \(\Y\), this likelihood model is implicitly assuming that all the information needed to estimate \(\theta_{d}\) is contained in the observed data. 
Should there be any error in the assumed functional form between \(\Y\) and the scaling factors \(\mathbf{S}_{dn}\) that translate the scale of \(\Y\) to the scale of \(\W\), then unacknowledged bias will result. 

\subsection{Pitfall 3: Singular Implicit Priors in Compositional Models}
Especially in the analysis of sequence count data, it has become increasingly popular to report results in terms of Centered Log-Ratio (CLR) coordinates. Misinterpretation of those results can be thought of as involving an implicit singular (i.e., zero variance) prior that again leads to unacknowledged bias. As an example, we discuss this phenomenon in regards to the popular ALDEx2 model for differential abundance analysis from sequence count data~\citep{fernandes2014unifying}. 

ALDEx2 is a model best described procedurally in three steps. In the first step, each multivariate observation \(\Y_{\cdot n}\) is modeled using an independent Bayesian Multinomial-Dirichlet model:
\begin{align}
  \Y_{\cdot n} &\sim \text{Multinomial}(\WC_{\cdot n}) \\
  \WC_{\cdot n} &\sim \text{Dirichlet}(\alpha). 
\end{align}
In the second step, posterior samples of \(\WC_{\cdot n}\) are normalized using the Centered Log-Ratio (CLR) transform
\[\eta_{\cdot n}=CLR(\WC_{\cdot n})=\left( \log \frac{\WC_{1n}}{G(\WC_{\cdot n})}, \dots \log\frac{\WC_{Dn}}{G(\WC_{\cdot n})} \right).\]
In the third step, for each transformed posterior sample and each \(d\in \{1, \dots, D\}\), ALDEx2 estimates the log-fold change in CLR transformed values as
\begin{equation} \label{eq:lfc-clr}
  \hat{\theta}_{d}= \mean_{x_{n}=1} \eta_{dn} - \mean_{x_{n}=0} \eta_{dn}. 
\end{equation}
For rhetorical simplicity, we present ALDEx2 as a fully Bayesian model. The
actual ALDEx2 model goes on, for each posterior sample of
\(\hat{\theta}_{d}\), to test the null hypothesis that \(\hat{\theta}_{d}=0\) using
a t-test or Wald test and then reports the mean of the resulting test
statistics and p-value.

The problem is not with ALDEx2 itself. Instead, the problem is that many researchers mistake estimates of \(\hat{\theta}_{d}\) for estimates \(\theta_{d}\): for example, they assume that if \(\hat{\theta}_{d}\) is positive, it suggests that the observed data supports the conclusion that the abundance of taxon \(d\) is higher in condition \(1\) than in condition \(0\). Yet this is not what is being estimated. If \(\hat{\theta}_{d}\) is positive, it suggests that the observed data supports the conclusion that the \textit{CLR transformed abundance} of taxon \(d\) is higher in condition \(1\) versus condition \(0\). In fact, the field of Compositional Data Analysis has long warned against this misinterpretation~\citep[pg. 32]{pawlowsky2015book}. To frame this as an implicit prior, consider that this misinterpretation is tantamount to assuming  \(\hat{\theta}_{d}=\theta_{d}\). Using the definition of the CLR transform and Equation 2 in the main text, this assumption can be written as:
\begin{align*}
\theta_{d}&=\hat{\theta}_{d} \\
\thetaC_{d}+\thetaT&=\thetaC_{d}+\hat{\thetaT} \\ 
\thetaT&=\hat{\thetaT} = -\mean_{n:x_{n}=1} \log G(\WC_{\cdot n}) +\mean_{n:x_{n}=0} \log G(\WC_{\cdot n})
\end{align*}
where \(\theta^{\tiny \mbox{tot}}\) and \(\theta^{\tiny \mbox{tot}}_{d}\) are defined as in Equation 2 in the main text. This can be written as the following implicit prior 
\[\thetaT \sim \delta[-\mean_{n:x_{n}=1} \log G(\WC_{\cdot n}) +\mean_{n:x_{n}=0} \log G(\WC_{\cdot n})]\]
where \(\delta[x]\) denotes a point mass centered at \(x\). 


\section{P-Value Calculation in ALDEx2}
\label{sec:aldex2-pvalues}

ALDEx2 conducts hypothesis tests on every Monte Carlo sample for every entity and computes a two-sided p-value for these tests. For a given entity, these p-values are averaged over the Monte Carlo samples, and a single p-value is returned. This type of averaging can be problematic since the p-values used are two-sided. To see this concretely, consider a hypothetical example with one taxtaxon, two Monte Carlo replicates, and 10 samples in each condition ($n_1 = n_2 =10$). Suppose the obtained two-sided test statistics are -3 and 3. Assuming $n_1 + n_2 - 2 = 18$ degrees of freedom, the corresponding two-sided p-values are 0.008 and 0.008, resulting in a reported p-value of 0.008. While both test results suggest there is an effect, they disagree about the direction of the effect, and simple aggregation of two sided p-values is misleading \citep{oosterhoff1969combination}. In the original ALDEx2 and Relaxed SSRV models, we expect that this will rarely, if ever, cause problems, but we warn that this may be problematic if more complex scale models are used and urge the user to use caution.

\section{Example of Scale Reliant Inference in Ecology}


A common task in ecology is to estimate the rate at which an animal species abundance changes
between two time-points \citep{udevitz1992change}. To avoid needlessly complicating this example, we consider only a
single species whose abundance is denoted at time \(t\) by \(\W_{t}\). In this case, the relative
abundance at time \(t\) is trivial \(\WC_{t}=1\) implying that \(\W_{t}=\WT_{t}\).
Ecologists often want to estimate the value of the logarithmic rate of change:
\begin{align*}
r&=\log \left( \frac{\W_{t+1}}{\W_{t}} \right)\\
&= \log \left( \frac{\WT_{t+1}}{\WT_{t}} \right).
\end{align*}
It is clear that this target estimand is fully scale reliant.

Yet many ecological surveys that attempt to estimate changes in species's abundance over time are
limited to using observational data where the scale, e.g., \(\W_{t}\), cannot be directly
measured. Instead, ecologists often use some variation on mark-recapture studies
\citep{krebs1999book}. In a mark-recapture study, animals are captured and released twice. During
the first capture, animals are marked prior to release, and the total number of animals captured are recorded (\(\Y^{1}\)). The second capture not only records the total number of captured
animals \(\Y^{2t}\) but also the number of those captured that had marks indicating that they had
been recaptured \(\Y^{2m}\). By performing independent mark-recapture studies both at time-point
\(t\) and \(t+1\), ecologists hope to collect sufficient information to estimate \(r\). We can
think of these data as being derived from a compositional survey, albeit a trivial compositional
survey since the \textit{relative abundance} in this simple 1 species example is trivially
identifiable as it is always equal to 1. Below, we illustrate a use of mark-recapture coupled with
a technique called the change-in-ratios method \citep{udevitz1992change} that is sometimes used to
estimate \(r\).

Suppose two separate mark-recapture experiments are undertaken at times $t$ and $t+1$. Let
$\{\Y^{1}_t, \Y^{2m}_t, \Y^{2t}_t\}$ denote the number of initially-marked individuals, the number of
recaptured marked individuals, and the total number of observed individuals during the recapture
survey at time $t$. We denote an analogous vector $\{\Y^{1}_{t+1}, \Y^{2m}_{t+1}, \Y^{2t}_{t+1}\}$
for time $t+1$. These data are often used to estimate the population size (species abundance) at
each time-point:
\begin{align*}
\hat{\W}_{t} &= \frac{\Y^{1}_{t}\Y^{2t}_{t}}{\Y^{2m}_{t}} \\ 
\hat{\W}_{t+1} &= \frac{\Y^{1}_{t+1}\Y^{2t}_{t+1}}{\Y^{2m}_{t+1}}
\end{align*}
In the change-of-ratios methods, these estimates are then used to estimate \(r\) as:
\begin{align*}
\hat{r} &= \log \left( \frac{\hat{\W}_{t+1}}{\hat{\W}_{t}} \right) \\ 
&= \log \left( \frac{\Y^{1}_{t+1}}{\Y^1_{t}} \right) + \log \left( \frac{\Y^{2t}_{t+1}/\Y^{2m}_{t+1}}{\Y^{2t}_t/\Y^{2m}_t} \right). 
\end{align*}
This is an applied estimator for \(r\).

The framework of SRI provides useful concepts to help catalyze discussions about this type of
analysis. Does this constitute fully scale reliant, partially scale reliant, or scale invariant
inference? Of course, as we have not explicitly specified a measurement model, it is not possible
to make any formal claim. Still, thinking through the mechanics of a mark-recapture study, it seems
reasonable to assume that these studies do not identify the system scale. Consider that
implicit in the above applied estimator is an assumption of homogeneous mixing and random
sampling: there is an assumption that the marked animals homogeneously mix into the population
such that the recapture probability \(\Y^{2m}_{t}/\Y^{2n}_{t}\) is equal to the proportion of marked
animals in the entire population. This assumption is clearly not exactly true, leaving fundamental
uncertainty in the system scale despite the observed data. It therefore seems reasonable to assume
that this is a partial compositional survey. In contrast, we expect that the degree of uncertainty
is likely far less than what is present in sequence count data, and it is likely unreasonable to
model these data as coming from a fully compositional survey. In either case, given that the
target estimand is one-to-one with the ratio of scales \(\WT_{t+1}/\WT_{t}\), it seems
clear that inference of \(r\) must be scale reliant. We therefore expect that strict Frequentist
consistency is likely not possible.

Overall, this discussion is merely meant to illustrate how the language of SRI can be applied to
different fields. Further work is needed if SRI is to provide a comprehensive framework for
understanding the role of scale in mark-recapture based ecological studies. 

\section{Data Analysis Details}
\subsection{Simulation Details and Results}
\label{sec:simulation-study}

The true abundances ($\W$) of 20 microbial taxa were simulated for $N=100$ samples equally split between a pre-antibiotic (\(x_{n}=0\)) and a post-antibiotic (\(x_{n}=1\)) condition using a Poisson model:
\begin{equation*}
\W_{dn} \sim
    \begin{cases}
        \mathrm{Poisson}(\lambda_{d, 0}) & \mathrm{if \, \, x_n = 0}.\\
        \mathrm{Poisson}(\lambda_{d, 1}) & \mathrm{if \, \, x_n = 1}.
    \end{cases}
\end{equation*}
For 16 of 20 taxa, we specified $\lambda_{d, 0} = \lambda_{d, 1}$ (not differentially abundant). For those taxa, 4 had \(\lambda_{d}=4,000\), 3 had \(\lambda_{d}=500\), and 9 had \(\lambda_{d}=400\). For the four taxa that were changing, all were decreased in the post-antibiotic condition compared to the pre-antibiotic condition: $\lambda_{(3,4,15,\text{ and } 20); 0} = \{4,000, 4,000, 400, 400\}$ versus $\lambda_{(3,4,15,\text{ and }20); 1} = \{3,000, 2,000, 200, 100\}$. The true abundances were re-sampled using multinomial re-sampling to reflect the loss of information due to sequencing:
\begin{equation*}
\Y_{\cdot n} \sim \mathrm{Multinomial}\left(M, \frac{\W_{\cdot n}}{\sum_{d=1}^D\W_{dn}}\right)\
\end{equation*}
Sequencing depth ($M$) was set to 5,000, and, unless otherwise
specified, sample size was set to 100, equally split between the two conditions ($n_{\text{pre}} = n_{\text{post}} = 50$). 

ALDEx2 and DESeq2 were fit using their respective Bioconductor
packages. For ALDEx2, 1,000 Monte Carlo samples
and a Welch's t-test were used. Taxa were called differentially abundant if their
Benjamini-Hochberg corrected p-values were less than 0.05. For DESeq2, taxa were
called differentially abundant if their Benjamini-Hochberg corrected p-values
based on the Wald test were less than 0.05. Our ALDEx2-based scale simulation random variable used the same conceptual approach as ALDEx2, replacing the normalization
with stochastic version of the CLR
(\(\log \WT_{n} \sim N(-\mean \left(\log \WC_{\cdot n}\right), \alpha^{2})\)). For the \textit{Relaxed} model, we set $\alpha = 0.5$.
Like with the original ALDEx2, we used a p-value cut off of 0.05 based on the
Benajamini-Hochberg corrected p-values after applying a t-test. 1,000 Monte Carlo
replicates were used.

\begin{figure}[H]
	\centering
	\includegraphics[width=4.5in]{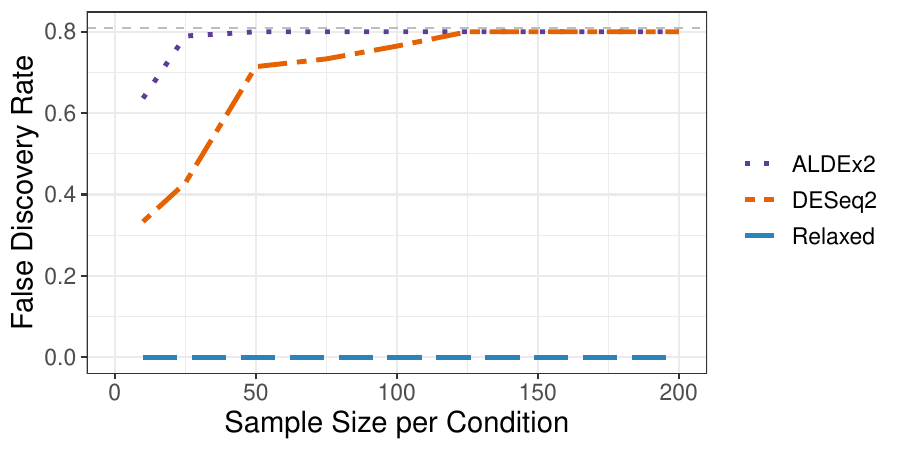}
	\caption{\footnotesize False discovery rate versus sample size for ALDEx2, DESeq2, and the \textit{Relaxed} SSRV. Data were simulated according to the procedure outlined in Section \ref{sec:simulation-study}, but the total sample size per condition was allowed to vary. For each
    method besides the SSRV, the false discovery rate increases even in the presence of infinite data. A false
    discovery rate of $0.80$ corresponds to all taxa being returned as differentially abundant as four of 20 taxa were truly changing. All methods were fit as outlined in Section \ref{sec:simulation-study}.}
	\label{fig:unack_bias}
\end{figure}

\subsection{Artificial Gut Application Details and Results}

\subsubsection{Bioreactor Culturing}

The bioreactor experiment consisted of 8 continuous-flow vessels (Multifors 2,
Infors) cultured over 25 total days according to previously established
protocols \citep{silverman2018dynamic, letourneau2021ecological}. Briefly,
vessels were inoculated with identical inoculum derived from a single healthy
stool donor. Vessels were maintained anaerobic under nitrogen, kept at 37
$^{\circ}$C, continuously stirred at 100 rpm using magnetic impeller
stir-shafts, and the total volume of media (400 mL modified Gifu Anaerobic
Medium) was turned over once per 24 hours. Samples for sequencing and flow
cytometry were collected daily at 4pm by syringe via a sterilized sampling port.
Since the volume of each vessel remained fixed at 400 mL and the vessels were
continuously mixed, the concentration of bacterial cells should be directly proportional to total bacterial load.

\subsubsection{Cell Quantification by Flow Cytometry}

Samples were serially diluted ten-fold for a final dilution of 1:1000. Serial dilutions were
performed in PBS, with the final dilution performed into wells of a 96-well plate containing PBS
with 1 $\mu$L SYBR green/DMSO (1:499 SYBR:DMSO) per 180 $\mu$L total volume. The final 200 $\mu$L
mixture including sample and SYBR green was incubated 15 min at 37 $^{\circ}$C
\citep{vandeputte2017quantitative}.

Flow cytometry was performed using a MACSQuant Flow Cytometer (Miltenyi Biotec). The 96-well plate
containing samples was maintained on a chill rack for the duration of the run. 10 $\mu$L was taken
for measurement. Forward scatter, side scatter, and SYBR fluorescence (using channel B1, log5, 274
V) were recorded.

Flow cytometry data was analyzed in FlowJo using a two-step gating strategy. A single sample
containing cells but no SYBR was used to develop gates which were then applied to all samples.
First, using a graph of forward scatter vs. side scatter, we roughly selected for the center of
point density, omitting extreme values (i.e. possibly representing cellular debris and/or particle
aggregates). Second, examining the SYBR channel, we excluded counts that fell within the range of
the no-SYBR control.

\subsubsection{16S rRNA-Gene Encoding Amplicon Sequencing}

16S rRNA gene amplicon sequencing was performed as previously described using custom barcoded
primers targeting the V4 region of the gene, resulting in 150 bp paired-end reads sequenced by
Illumina MiniSeq according to previously published protocols \citep{silverman2018dynamic,
  letourneau2021ecological}. DADA2 was used to identify and quantify sequence variants (SVs) in our
dataset, using version 123 of the Silva database. Taxa that had counts of less than 2,500 across all
samples were filtered and amalgamated into a separate category.

\subsubsection{Data Analysis}
\begin{figure}[H]
   \centering \includegraphics[width=3in]{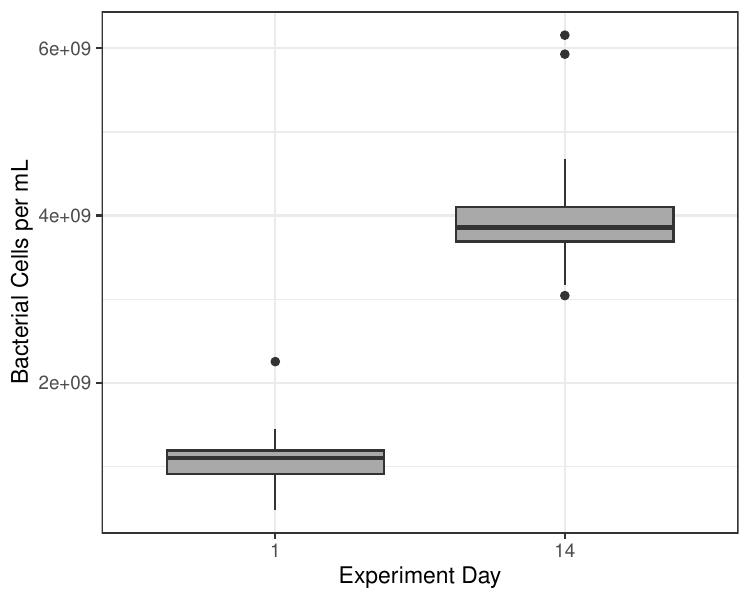}
   \caption{Variation of flow cytometry data by day. Note that there was a higher concentration of
     cells on day 14 regardless of vessel. This has biological validity based on the experimental
     goals of the study.}
   \label{supp-fig:flow}
 \end{figure}

For our analysis, we restricted ourselves to days 1 and 14. Furthermore, we used data from vessels
3, 4, 5, 6, 7, and 8 only due to an experimental error that occurred in Vessels 1 and 2 that causing
the volume of liquid in the vessels to not be maintained during the experiment. We noted a
difference between flow cytometry concentrations between days regardless of vessel. See
Figure~\ref{supp-fig:flow} for details.
 
We tested a model of differential expression including indicator terms for each vessel and day. For
ALDEx2, we used the generalized linear model test with day and vessel as the predictor variables. 1,000 Monte Carlo samples were drawn. We assessed for
differentially abundant taxa based on the p-values ($\alpha = 0.05$) for day alone. For DESeq2, we
employed the likelihood ratio test with a reduced model containing only the vessel and the full model containing both vessel and day. We assessed for
differential abundance using the same criteria as ALDEx2.

For the SSRV based on the Multinomial Log Normal model, we set our prior parameters as follows: $\nu = D+3$
and $\Xi=I_D$ reflecting our weak prior knowledge that the taxa were independent;
$\Gamma = \text{BlockDiagonal}(10I_{6}, 1)$ reflecting our assumption that the some taxa were more
abundant than others at baseline and assuming that the signal-to-noise ratio with respect to
differential abundance was most likely on the order of 1 \citep{silverman2018dynamic}; and finally
$M=0$ which reflected that we did not have \textit{a priori} knowledge differentiating the taxa.

Using results presented in Supplementary Section~\ref{supp:scale-pim}, we can interpret the scale model implied by this prior choice. Using estimates of $\W$ obtained from the PIM model, we estimated $\thetaT = -0.46$ (95\% credible interval: $[-1.05, 0.12]$) under this prior specification. Thus, these prior choices equate to the belief that there is a small decrease in the total microbial load on day 14 compared to day 1, an assumption at odds with the flow cytometry measurements ($\thetaT = 1.22$).

\begin{figure}[ht]
  \centering
  \includegraphics[width=4in]{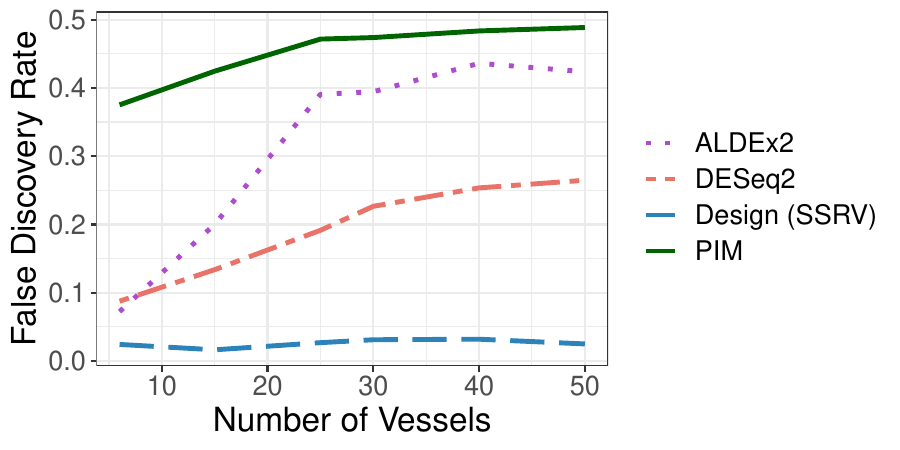}
  \caption{\footnotesize False discovery rates (FDR)  by method (ALDEx2, DESeq2, SSRV - Design, PIM) with increasing sample sizes. Additional samples were drawn using bootstrap resampling. FDR increases with increasing sample sizes for three of the methods (ALDEx2, DESeq2, PIM), demonstrating unacknowledged bias and highlighting the need for careful selection of priors in the case of the PIM. In contrast, a SSRV with a  carefully designed scale model (SSRV (Design)) has low FDR asymptotically. Reported results are averages over 25 re-sampling experiments for each sample size.}
  \label{fig:unacknowledged-bias-real-data}
\end{figure}

31 (all) different sequence variants were declared significant by at least one method. Taxonomic information for these significant sequence variants is presented in Table 2. All results are presented in the main text except those presented in Figure \ref{fig:unacknowledged-bias-real-data}. Here, we used bootstrap resampling to resample the number of vessels per condition ($N \in [6, 50]$ vessels per condition). The Gold Standard model was fit for each sample, and several models were benchmarked off of its results: ALDEx2, DESeq2, the Design SSRV, and the Bayesian PIM discussed in the main text. The bootstrap resampling and model fitting were repeated 25 times for each sample size. Due to the large increase in scale between day 1 and day 14, a result was recorded as a true positive if it was significant and the sign of the estimated effect matched between a given model and the Gold Standard model. Thus, both results that were significant in a given model but not in the Gold Standard model and were significant in both but the models disagreed on sign were recorded as false positives. Average false discovery rates (FDR) by method by sample size are reported in Figure \ref{fig:unacknowledged-bias-real-data}. As the number of vessels increases, ALDEx2, DESeq2, and the Bayesian PIM display increasing FDR, a hallmark of unacknowledged bias. In the case of the Bayesian PIM, even though uncertainty in scale was incorporated into the model, poor prior specification dominated, leading to elevated FDR and highlighting the need for correct prior specification. Conversely, the Design-based SSRV controlled FDR rates asymptotically by using a carefully-chosen scale model.

 \begin{spacing}{0.8}
 \begin{table}[H]
    \label{tab:otuLabels}
    \caption{Taxonomic information of significant sequence variants identified in the Real Data Analysis.}
     \begin{center}
       \begin{tabular}{ c c } 
         \textbf{Sequence}  & \textbf{Taxonomic Information}\\
         \hline
         \hline
         1  & \textit{Bacteroides spp.} \\ 
         2  & \textit{Escherichia/Shigella spp.}\\
         3  & \textit{Bacteroides uniformis}\\ 
         4  & \textit{Peptoniphilus spp.} \\ 
         5  &  \textit{Pyramidobacter piscolens}\\
         6  & \textit{Enterococcus spp.}\\
         7  & \textit{Veillonella spp.} \\
         8  & \textit{Bacteroides spp.}\\
         9  &  \textit{Streptococcus spp.} \\
         10 & \textit{Anaerosalibacter bizertensis}\\
         11 & \textit{Bacteroides stercoris} \\
         12 & \textit{Bacteroides caccae}\\
         13 & \textit{Megasphaera spp.}\\
         14 & \textit{Peptoniphilus lacrimalis}\\
         15 & \textit{Lachnoclostridium spp.}\\
         16 & \textit{Dorea formicigenerans} \\
         17 & \textit{Phascolarctobacterium faecium}\\
         18 & \textit{Veillonella spp.}\\
         19 & \textit{Lachnoclostridium spp.}\\
         20 & \textit{Clostridium sensu stricto spp.}\\
         21 & \textit{Veillonella spp.} \\
         22 & \textit{Peptoclostridium difficile}\\
         23 & \textit{Erysipelatoclostridium ramosum}\\
         24 & \textit{Hungatella spp.}\\
         25 & \textit{Flavonifractor plautii }\\
         26 & \textit{Bifidobacterium spp.}\\
         27 & \textit{Parabacteroides merdae}\\
         28 & \textit{Anaerococcus vaginalis}\\ 
         29 & \textit{Blautia spp.}\\
         30 & \textit{Eubacterium spp.}\\
         31 & \textit{Anaerostipes hadrus}\\
         \hline
         \hline
    \end{tabular}
    \end{center}
\end{table}
\end{spacing}

\section{Deriving the Multinomial Logistic Normal Model as a SSRV}

In this section, we show how the multinomial logistic-normal model for sequence count data presented in the manuscript can be inferred as a SSRV.

\subsection{Derivation of Measurement Model and Scale Model}

For a $D \times N$ count matrix $\Y$, the multinomial logistic normal can be written as:

\begin{align}
  \label{eq:target}
  \Y_{\cdot n} &\sim \mathrm{Mult}(\WC_{\cdot n})\\
  \W_{\cdot n} &= \WT_j \WC_{\cdot n}\\
  \W_{\cdot n} &= \exp(\psi_{\cdot n})\\
  \psi_{\cdot n} &\sim N(\B X_{\cdot n}, \OmegaBF, \mathbf{I})\\
  \B &\sim N(\M, \OmegaBF, \GammaBF)\\
  \OmegaBF &\sim IW(\nu, \XiBF)
           \label{eq:target-end}
\end{align}
where $n = 1,...,N$ denotes the index of the sample. First, we define $\G$ as a linear map:

\begin{equation}
  \G = \begin{bmatrix}
    \F\\
    H\\
  \end{bmatrix}
\end{equation}
where $\F = [\mathbf{I}_{D-1}, -1_{D-1}]$ and $H = \mathbf{1}_D$. Note that, when applied to a vector or matrix on the log scale, $\F$ corresponds to the additive logratio (ALR) transformation whereas $\G$ is orthogonal to $\F$ and represents the remaining required information to reconstruct $\W$. This differs from the notion of total presented in the main text.

For any arbitrary vector
$\theta$ and matrix $\SigmaBF$, we have:

\begin{align}
  \G \theta &= \begin{bmatrix} \F\theta\\ H\theta\\ \end{bmatrix} = \begin{bmatrix} \theta^\dagger \\ \theta^\perp\\ \end{bmatrix}\\
  \G\Sigma &= \G \SigmaBF \G^T = \begin{bmatrix} \F\SigmaBF \F^T & \F\SigmaBF H^T\\ H \SigmaBF \F^T & H \SigmaBF H^T\\ \end{bmatrix} = \begin{bmatrix} \SigmaBF^{\dagger} & {\Sigma^+}^T\\ \Sigma^+ & \Sigma^{\perp}\\ \end{bmatrix}
\end{align}
where the superscript $\dagger$ denotes the transformation induced by $\F$, and the superscript $\perp$ denotes the transformation induced by $H$. This purposefully differs from the notation in the main text to highlight the fact that these two are not directly analogous to the composition and total.

Applying $\G$ to $\psiBF$ from the model defined in Equations \ref{eq:target}-\ref{eq:target-end}:
\begin{equation}
  \G \psiBF = \begin{bmatrix} \psiBF^{\dagger}\\ \psi^{\perp}\\ \end{bmatrix}\\
\end{equation}
Since it is an affine transformation, we get (Theorem 2.3.10, \cite{gupta2000matrix}, page 64):
\begin{align}
  \begin{bmatrix} \psiBF^{\dagger}\\ \psi^{\perp}\\ \end{bmatrix} &\sim N(\G\B X, \G\OmegaBF \G^T, \mathbf{I})\\
  \begin{bmatrix} \psiBF^{\dagger}\\ \psi^{\perp}\\ \end{bmatrix} &\sim N\left(\begin{bmatrix} \B^{\dagger}\\ B^{\perp}\\ \end{bmatrix} X, \begin{bmatrix} \OmegaBF^{\dagger} & {\Omega^+}^T\\ \Omega^+ & \Omega^{\perp}\\ \end{bmatrix} , \mathbf{I}\right)\\
  \begin{bmatrix} \psiBF^{\dagger}\\ \psi^{\perp}\\ \end{bmatrix} &\sim N\left(\begin{bmatrix} \B^{\dagger}\\ B^{\perp}\\ \end{bmatrix} X,\OmegaBF^*, \mathbf{I}\right)
  \label{eq:psi-model}
\end{align}
from the properties discussed before. Now, we need to marginalize out $\B$ and $\OmegaBF$. Starting
with $\B$, we can write Equation \ref{eq:psi-model} as:
\begin{equation}
  \begin{bmatrix} \psiBF^{\dagger}\\ \psi^{\perp}\\ \end{bmatrix} = \begin{bmatrix} \B^{\dagger}\\ B^{\perp}\\ \end{bmatrix} \X + \mathbf{E}^B, \; \mathbf{E}^B \sim N(\mathbf{0}, \OmegaBF^*, \mathbf{I})
\end{equation}
Similarly, since $B\sim N(\M, \OmegaBF, \GammaBF)$, by Theorem 2.3.10 (\cite{gupta2000matrix}, page 64), we
get:
\begin{equation}
  \G \B = \begin{bmatrix} \B^{\dagger}\\ B^{\perp}\\ \end{bmatrix} \sim N\left(\begin{bmatrix} \M^{\dagger}\\ M^{\perp}\\ \end{bmatrix}, \OmegaBF^*, \GammaBF\right)
\end{equation}
Then, we can write:
\begin{equation}
  \begin{bmatrix} \B^{\dagger}\\ B^{\perp}\\ \end{bmatrix} = \begin{bmatrix} \M^{\dagger}\\ M^{\perp}\\ \end{bmatrix} + \mathbf{E}^*, \; \mathbf{E}^* \sim N(0,\OmegaBF^*, \GammaBF).
\end{equation}
This implies that
\begin{align}
  \begin{bmatrix} \psiBF^{\dagger}\\ \psi^{\perp}\\ \end{bmatrix} &= \left(\begin{bmatrix} \M^{\dagger}\\ M^{\perp}\\ \end{bmatrix} + \mathbf{E}^*\right)\X + \mathbf{E}^B\\
                                                                  &= \begin{bmatrix} \M^{\dagger}\\ M^{\perp}\\ \end{bmatrix}\X +\mathbf{E}^*X + \mathbf{E}^B
\end{align}
where $\mathbf{E}^B$ and $\mathbf{E}^*$ are defined as before. Using Theorem 2.3.10 (\cite{gupta2000matrix}, page 64),
the distribution of $\mathbf{E}^* \X$ is:
\begin{equation}
  \mathbf{E}^* \X \sim N(\mathbf{0}, \OmegaBF^*, \X^T \GammaBF \X)
\end{equation}
Then, by \cite{gupta2000matrix} (page 82), we get
\begin{equation}
  \begin{bmatrix} \psiBF^{\dagger}\\ \psi^{\perp}\\ \end{bmatrix} \sim N\left(\begin{bmatrix} \M^{\dagger}\\ M^{\perp}\\ \end{bmatrix}\X, \OmegaBF^*, \mathbf{I} + \X^T \GammaBF \X\right)
\end{equation}
To marginalize out $\OmegaBF^*$, we start by applying $\G$ to $\OmegaBF$:
\begin{equation}
  \G\OmegaBF \G^T= \OmegaBF^* \sim \mathrm{IW}(\nu,\XiBF^*)
\end{equation}
where
$\XiBF^* = \G \XiBF \G^T = \begin{bmatrix} \XiBF^{\dagger} & {\Xi^+}^T\\ \Xi^+ & \Xi^{\perp}\\ \end{bmatrix}$. Thus, from the definition of the matrix-t distribution \citep{silverman2022bayesian}, we get
\begin{equation}
  \begin{bmatrix} \psiBF^{\dagger}\\ \psi^{\perp}\\ \end{bmatrix} \sim T\left( \nu, \begin{bmatrix} \M^{\dagger}\\ M^{\perp}\\ \end{bmatrix}\X, \XiBF^*, \mathbf{I}+\X^T \GammaBF \X\right).
\end{equation}
From \cite{gupta2000matrix} (page 163), the marginal distribution of $\psiBF^{\dagger}$ is:
\begin{equation}
  \psiBF^{\dagger} \sim T(\nu, \M^{\dagger} \X, \XiBF^{\dagger}, \mathbf{I}+\X^T \GammaBF \X)
\end{equation}
which implies the model for $\psiBF^{\dagger}|Y$ is:
\begin{align}
  \Y_{\cdot n} &\sim \mathrm{Mult}(\WC_{\cdot n})\\
  \psiBF^{\dagger} &\sim \mathrm{ALR}_D(\WC)\\
  \psiBF^{\dagger} &\sim T(\nu, \M^{\dagger} \X, \XiBF^{\dagger}, \mathbf{I}+\X^T \GammaBF \X).
\end{align}
where $ALR(\cdot)$ denotes the additive log-ratio transform. Note that this is exactly the collapse
sampler from \cite{silverman2022bayesian}.

\subsection{Derivation of Implied Scale Model for Bayesian PIM}
\label{supp:scale-pim}

Recall from the previous section:
\begin{equation}
  \begin{bmatrix} \psiBF^{\dagger}\\ \psi^{\perp}\\ \end{bmatrix} \sim T\left( \nu, \begin{bmatrix} \M^{\dagger}\\ M^{\perp}\\ \end{bmatrix}\X, \XiBF^*, \mathbf{I}+\X^T \GammaBF \X\right).
\end{equation}
Then, from \cite{gupta2000matrix} (page 138), the distribution of $\psi^{\perp}| \psiBF^{\dagger}$
is a multivariate T-distribution:

\begin{align}
  \psi^{\perp} | \psiBF^{\dagger} &\sim T[\nu + D, M^{\perp} \X + (\Xi^+)^T (\XiBF^{\dagger})^{-1} (\psiBF^{\dagger} - \M^{\dagger} \X),\\
                                  & \Xi_{22 \cdot 1}, (\mathbf{I}+\X^T \GammaBF \X)(I + (\mathbf{I}+\X^T \GammaBF \X)^{-1}(\psiBF^{\dagger} - \M^{\dagger} \X)^{T} (\XiBF^{\dagger})^{-1}(\psiBF^{\dagger} - \M^{\dagger} \X))]
\end{align}
where $\Xi_{22 \cdot 1} = \Xi^{\perp} - (\Xi^+) (\XiBF^{\dagger})^{-1} (\Xi^+)^T$.

\subsection{Collapse-Uncollapse Sampling Details}
\label{supp-collapse}
While conceptually straightforward, sampling SSRVs requires efficient sampling for two separate
stages: a measurement model and a scale model. These samples are combined to calculate the target estimand. For the case of the
multinomial logistic normal model, our sampling details closely mimic the Collapse-Uncollapse (CU)
sampler of \cite{silverman2022bayesian}. Intuitively, the basic sampler follows three steps:

\begin{enumerate}
  \item Sample from the partially identified measurement model $P(\WC|Y)$.
  \item Sample from the scale model $P(\WT | \WC)$.
  \item Compute the target estimand from the samples of $\WC$ and $\WT$.
\end{enumerate}

Sampling from the scale model is typically efficient as it is low-dimension and flexible. To date,
efficient routines for these models are already implemented in \texttt{R}. However, sampling from Step 1 is cumbersome but was recently made much more efficient and
scalable by \cite{silverman2022bayesian}. Their approach relies on rewriting the above multinomial
logistic-normal model as a marginally latent matrix-t process. Then, they develop efficient
inference for this class of models using a Collapse-Uncollapse (CU) sampling approach. Such an
approach vastly improves computational efficiency, increasing the usability of these models to much
larger sequence count data sets. See \cite{silverman2022bayesian} for complete details on
implementation.

We utilize the logic behind the CU sampler to infer Step 1. These routines are efficiently
implemented in the R package \texttt{fido} \citep{silverman2022bayesian}. Formal
implementation details for scale simulation are outlined in Algorithm~\ref{alg:collapse}.

\begin{algorithm}[H]
  \label{alg:collapse}
  \SetAlgoLined \SetKwBlock{DoParallel}{do in parallel}{end} \KwData{Y} \KwResult{S samples of $\theta$}
  Sample $\{\WC[(1)], ... , \WC[(S)] \} \sim P(\WC|\Y)$ using the Collapse sampler of \cite{silverman2022bayesian};\\
  Sample $\{\WT[(1)], ... , \WT[(S)] \} \sim P(\WT|\WC)$;\\
    \For{$i=1,...,S$}{Calculate the value of the target estimand $\theta^{(i)}$ using $\W^{(i)} = \WC[(i)] \WT[(i)]$}
  \caption{Modified Collapse-Uncollapse Sampler for Scale Simulation}
\end{algorithm}

\newpage
\bibliographystyle{apalike} 
\bibliography{refs}